\documentclass[aps,prl,showpacs,twocolumn,superscriptaddress]{revtex4-2}

\usepackage{amsmath}
\usepackage{amsfonts}
\usepackage{amssymb}
\usepackage{epsfig}

\usepackage{braket}

\usepackage{graphicx}

\usepackage{bbold}

\usepackage{hyphenat}
\usepackage{hyperref}
\usepackage{placeins}	
\usepackage{color}

\usepackage{array}
\newcolumntype{C}[1]{>{\centering\arraybackslash}m{#1}}
\usepackage{overpic}

\renewcommand{\vec}{\mathbf}

\DeclareUnicodeCharacter{0301}{\'{e}}

\usepackage[normalem]{ulem}

\usepackage{booktabs}
\usepackage{diagbox}
\usepackage{siunitx}
\AtBeginDocument{
	\heavyrulewidth=.08em
	\lightrulewidth=.05em
	\cmidrulewidth=.03em
	\belowrulesep=.65ex
	\belowbottomsep=0pt
	\aboverulesep=.4ex
	\abovetopsep=0pt
	\cmidrulesep=\doublerulesep
	\cmidrulekern=.5em
	\defaultaddspace=.5em
}
\begin{document}

\title{Quantum critical phase transition between two topologically-ordered phases in the Ising toric code bilayer}
\author{Raymond Wiedmann}
\affiliation{Lehrstuhl f\"ur Theoretische Physik I, Staudtstra{\ss}e 7, Universit\"at Erlangen-N\"urnberg, D-91058 Erlangen, Germany}

\author{Lea Lenke}
\affiliation{Lehrstuhl f\"ur Theoretische Physik I, Staudtstra{\ss}e 7, Universit\"at Erlangen-N\"urnberg, D-91058 Erlangen, Germany}

\author{Matthias R. Walther}
\affiliation{Lehrstuhl f\"ur Theoretische Physik I, Staudtstra{\ss}e 7, Universit\"at Erlangen-N\"urnberg, D-91058 Erlangen, Germany}

\author{Matthias M\"uhlhauser}
\affiliation{Lehrstuhl f\"ur Theoretische Physik I, Staudtstra{\ss}e 7, Universit\"at Erlangen-N\"urnberg, D-91058 Erlangen, Germany}

\author{Kai Phillip Schmidt}
\affiliation{Lehrstuhl f\"ur Theoretische Physik I, Staudtstra{\ss}e 7, Universit\"at Erlangen-N\"urnberg, D-91058 Erlangen, Germany}

\begin{abstract}
We demonstrate that two toric code layers on the square lattice coupled by an Ising interaction display two distinct phases with intrinsic topological order. The second-order quantum phase transition between the weakly-coupled $\mathbb{Z}_2\times\mathbb{Z}_2$ and the strongly-coupled $\mathbb{Z}_2$ topological order can be described by the condensation of bosonic quasiparticles from both sides and belongs to the 3d Ising$^*$ universality class. This can be shown by an exact duality transformation to the transverse-field Ising model on the square lattice, which builds on the existence of an extensive number of local $\mathbb{Z}_2$ conserved parities. These conserved quantities correspond to the product of two adjacent star operators on different layers. Notably, we show that the low-energy effective model derived about the limit of large Ising coupling is given by an effective single-layer toric code in terms of the conserved quantities of the Ising toric code bilayer. The two topological phases are further characterized by the topological entanglement entropy which serves as a non-local order parameter.    
\end{abstract}

\maketitle

Quantum matter exhibiting intrinsic topological order \cite{Wen_1989,Wen_1990,Wen_2004} displays macroscopic entanglement of the ground state, fractional excitations with unconventional anyonic particle statistics 
\cite{Leinaas_1977,Wilczek_1982}, or fracton topological order in three dimensions \cite{Chamon_2005,Bravyi_2011,Haah_2011,Yoshida_2013,Vijay_2015,Vijay_2016}. These fascinating physical properties give rise to potential applications in quantum technologies in terms of topological quantum computation with non-Abelian anyons or quantum memories exploiting the topological order of the ground state \cite{Kitaev_2003,Nayak_2008}. Experimentally, topological order is investigated either in fractional quantum Hall systems \cite{Laughlin_1983,Tsui_1982} or frustrated quantum magnets \cite{Balents_2010,Jackeli_2010,Singh_2012,Plumb_2014,Banerjee_2017,Savary_2017,Banerjee_2018} in condensed matter physics as well as with quantum simulators utilizing trapped ions \cite{Han_2007}, photons \cite{Lu_2009,Pachos_2009} or NMR \cite{Du_2007,Feng_2012,Peng_2014} and is proposed for several other platforms \cite{Micheli_2006,Paredes_2008,Sameti_2017} in quantum optics.

One fundamental aspect in studying topological order is the understanding of phase transitions out of topologically ordered phases, since, contrary  to  conventional  phases,  such phases  can not  be  described  by local  order parameters so that Landau-Ginzburg theory is not applicable. Hence, an extended framework is needed. One promising approach to describe quantum phase transitions out of topological phases relies on the condensation of bosonic quasiparticles, also dubbed topological symmetry breaking \cite{Bais_2002,Bais_2007,Bais_2009,Burnell_2011,Burnell_2018}, which has been verified microscopically for a variety of models displaying phase transitions between topological and non-topological phases \cite{Trebst_2007,Hamma_2008_b,Yu_2008,Vidal_2009,Vidal_2011,Dusuel_2009,Tupitsyn_2010,Wu_2012,Dusuel_2011,Schmidt_2013,Jahromi_2013,Morampudi_2014,Schulz_2016,Zhang_2017,Vanderstraeten_2017}.

The most paradigmatic model in this context is the toric code \cite{Kitaev_2003}, which is an exactly solvable two-dimensional quantum spin model. Its ground state displays intrinsic $\mathbb{Z}_2$ topological order 
and elementary excitations with mutual Abelian statistics. The exact solvability of the toric code gave rise to many studies investigating the physical properties of intrinsic topological order like its quantum robustness \cite{Trebst_2007,Hamma_2008_b,Yu_2008,Vidal_2009,Vidal_2011,Dusuel_2009,Tupitsyn_2010,Wu_2012,Dusuel_2011,Schmidt_2013,Morampudi_2014,Zhang_2017,Vanderstraeten_2017},  the effects of thermal fluctuations \cite{Alicki_2009,Castelnovo_2007,Nussinov_2009_b}, the properties of entanglement measures \cite{Halasz_2012,Santra_2014} and dynamical  correlation functions \cite{Kamfor_2014}, or extensions to topological order in 3d \cite{Hamma_2005,Nussinov_2008,Reiss_2019}. In particular, the 
    breakdown of the toric code due to an external field serves as a standard model for a topological phase transition between a topological and a non-topological phase which can be described by the condensation of bosonic quasiparticles \cite{Trebst_2007,Hamma_2008_b,Yu_2008,Vidal_2009,Vidal_2011,Dusuel_2009,Tupitsyn_2010,Wu_2012,Dusuel_2011,Morampudi_2014,Zhang_2017,Vanderstraeten_2017}. 

\begin{figure}[t]
        \centering
        \includegraphics[width=\columnwidth]{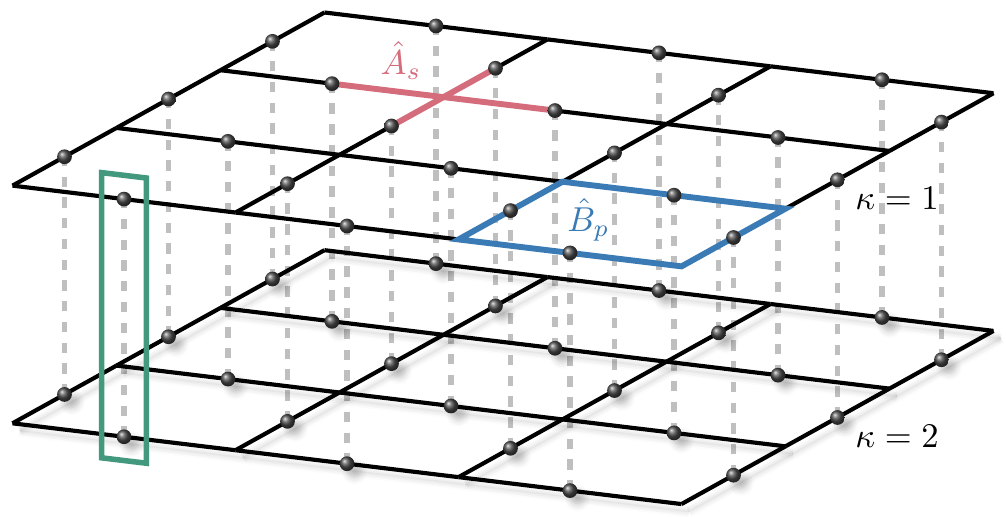}
        \caption{Sketch of the ITCB. Circles represent spin-1/2 degrees of freedom located on the links of the upper ($\kappa=1$) and lower ($\kappa=2$) toric code layer. Star and plaquette operators $\hat{A}_{s,\kappa}$ and $\hat{B}_{p,\kappa}$ are illustrated as red cross and blue plaquette, respectively. Thick dashed lines denote the inter-layer Ising interactions with coupling strength $I$. The rectangular box illustrates an Ising dimer.}
        \label{Fig::ITCB}
\end{figure}

In contrast, apart from bilayer fractional quantum Hall systems \cite{Wen_2010,Barkeshli_2010,Moeller_2014} and certain lattice models \cite{Bombin_2008,Morampudi_2014,Schulz_2016}, second-order phase transitions between two distinct topological phases are much less investigated and a paradigmatic example is still lacking. This is exactly the main focus of this letter. We demonstrate that the Ising toric code bilayer (ITCB) hosts the simplest kind of continuous quantum phase transition between two distinct phases possessing $\mathbb{Z}_2\times\mathbb{Z}_2$ and $\mathbb{Z}_2$ intrinsic topological order. The associated quantum phase transition can be described by the condensation of bosonic quasiparticles from both sides and it lies in the 3d Ising$^*$ universality class, which can be deduced from an exact duality mapping to the transverse-field Ising model on the square lattice.  

{\it{Model:}}
We study the ITCB as illustrated in Fig.~\ref{Fig::ITCB} where two toric code layers are coupled symmetrically by an Ising interaction. 
The Hamiltonian of the ITCB is given by
\begin{align}
 \mathcal{H}_{\rm ITCB}=-J\sum_{\kappa }\mathcal{H}_{\rm TC}^{(\kappa)}-I\sum_{i} \sigma_{i,1}^z\sigma_{i,2}^z\,, \label{eq:itcb}
\end{align}
where $\kappa=1$ ($\kappa=2$) denotes the upper (lower) layer, $\sigma_{i,\kappa}^\alpha$ represents the usual Pauli matrices with flavor \mbox{$\alpha\in\{x,z\}$} acting on the site in layer $\kappa$ of Ising dimer $i$, and $\mathcal{H}_{\rm TC}^{(\kappa)}$ is the Hamiltonian of the toric code   
\begin{align}
 \mathcal{H}_{\rm TC}^{(\kappa)}= \sum_s \hat{A}_{s,\kappa}+ \sum_p \hat{B}_{p,\kappa} \label{eq:tc}
\end{align}
with star and plaquette operators $\hat{A}_{s,\kappa}=\prod_{i\in s}\sigma_{i,\kappa}^x$ and $\hat{B}_{p,\kappa}=\prod_{i\in p}\sigma_{i,\kappa}^z$, respectively. These operators have eigenvalues $a_{s,\kappa},b_{p,\kappa}\in\{\pm 1\}$, which is a consequence of \mbox{$\hat{A}^2_{s,\kappa}=\hat{B}^2_{p,\kappa}=\mathbb{1}$}. In the following we focus on $J,I>0$, since all other cases can be mapped to this one, because the toric code and the Ising interaction have energy spectra which are symmetric under sign reversal.

Individual plaquette operators $\hat{B}_{p,\kappa}$ commute with $\mathcal{H}_{\rm ITCB}$ so that the $b_{p,\kappa}$ are conserved quantities. This is not the case for the eigenvalues $a_{s,\kappa}$ of the star operators $\hat{A}_{s,\kappa}$ whenever $I\neq 0$, because \mbox{$[\hat{A}_{s,\kappa},\sigma_{i,1}^z\sigma_{i,2}^z]\neq 0$} if $i\in s$. However, the ITCB still hosts a locally conserved parity quantum number $a_s^==\pm 1$ per star $s$. These parities are the eigenvalues of $\hat{A}_{s}^=\equiv \hat{A}_{s,1}\hat{A}_{s,2}$ for which $[\hat{A}_{s}^= ,\mathcal{H}_{\rm ITCB}]=0$ holds $\forall s$. As a consequence, the total Hilbert space of dimension $2^N$ with $N$ the number of spins decomposes exactly in decoupled subspaces of dimension $2^N/(2^{N/2}2^{N/4})=2^{N/4}$ with fixed $N/2$ quantum numbers $b_{p,\kappa}$ and $N/4$ parity quantum numbers $a_s^=$. 

{\it{Large-Ising phase:}} In the limit $J=0$, the ITCB reduces to isolated Ising dimers (see Fig.~\ref{Fig::ITCB}). Each Ising dimer has two degenerate ground states $\ket{\uparrow\uparrow}$, $\ket{\downarrow\downarrow}$ with energy $-I$ and two degenerate excited states $\ket{\uparrow\downarrow}$, $\ket{\downarrow\uparrow}$ with energy $+I$. The ground-state manifold is therefore extensively degenerate. 

In a next step we reformulate these four states by setting $\ket{\uparrow\uparrow} \equiv \ket{\Uparrow 0}$, $\ket{\downarrow\downarrow} \equiv \ket{\Downarrow 0}$,  $\ket{\uparrow\downarrow} \equiv \ket{\Uparrow 1}$, and  $\ket{\downarrow\uparrow} \equiv \ket{\Downarrow 1}$, where the first index refers to the states $\ket{\uparrow}$ and $\ket{\downarrow}$ of a pseudo-spin 1/2 $\vec{\tau}$ and the second index refers to the absence $\ket{0}$ (presence $\ket{1}$) of a hardcore boson representing a ferromagnetic (anti-ferromagnetic) dimer state \cite{Vidal_2008}. Introducing corresponding hardcore boson operators $\hat{b}^{\vphantom{\dagger}}_i$, $\hat{b}_i^\dagger$, we can express the Pauli matrices $\sigma^\alpha_{i,\kappa}$ as $\sigma_{i,1}^x = \tau_i^x (\hat{b}_i^\dagger + \hat{b}_i^{\vphantom{\dagger}})$, $\sigma_{i,2}^x = \hat{b}_i^\dagger + \hat{b}_i^{\vphantom{\dagger}}$, $
    \sigma_{i,1}^y = \tau_i^y (\hat{b}_i^\dagger + \hat{b}_i^{\vphantom{\dagger}})$, $\sigma_{i,2}^y = {\rm i} \tau_i^z (\hat{b}_i^\dagger - \hat{b}_i^{\vphantom{\dagger}})$, $\sigma_{i,1}^z = \tau_i^z$, and $\sigma_{i,2}^z = \tau_i^z (1 - 2 \hat{b}_i^\dagger \hat{b}_i^{\vphantom{\dagger}})$. 

The ITCB can then be rewritten exactly as follows
\begin{widetext}
\begin{align}\label{eq:H_BLTC_newparticles}
    \mathcal{H}_\mathrm{ITCB} = -\frac{IN}{2}+ 2I\sum_i \hat{n}_i &- J\sum_s (\mathbb{1}+\hat{\Tilde{A}}_s )  \prod_{i \in s} (\hat{b}_i^\dagger + \hat{b}_i^{\vphantom{\dagger}}) - J\sum_p \hat{\Tilde{B}}_p \left( 1 + \prod_{i \in p} (-1)^{\hat{n}_i}\right) 
\end{align}
\end{widetext}
with the local counting operators $\hat{n}_i=\hat{b}_i^\dagger \hat{b}_i^{\vphantom{\dagger}}$ and the effective pseudo-spin star and plaquette operators \mbox{$\hat{\Tilde{A}}_s \equiv \prod_{i \in s}\tau_i^x$} and $\hat{\Tilde{B}}_p \equiv \prod_{i \in p}\tau_i^z$. Interestingly, one finds 
\begin{align}
\hat{A}_{s}^= &= \hat{\Tilde{A}}_s\prod_{i \in s}(\hat{b}_i^\dagger  \hat{b}_i^{\vphantom{\dagger}}+ \hat{b}_i^{\vphantom{\dagger}}\hat{b}_i^\dagger)^2=\hat{\Tilde{A}}_s
\end{align}
so that the eigenvalues of the effective star operators are identical to 
$a_s^=$. One further has $\hat{B}_{p,1}=\hat{\Tilde{B}}_p$ and 
\begin{align}\label{Eq::Bp2}
 \hat{B}_{p,2}=\prod_{i \in p}(-1)^{\hat{n}_i}\hat{\Tilde{B}}_p\quad .
\end{align}
The prefactor $\prod_{i \in p}(-1)^{\hat{n}_i}$ is $+1$ ($-1$) if an even (odd) number of bosons are present on plaquette $p$. It is further apparent that each (local) operator in $\mathcal{H}_\mathrm{ITCB}$ changes the total number of bosons always by an even number so that the parity $\prod_{i \in p}(-1)^{\hat{n}_i}$ is conserved locally. 

It is now possible to derive an effective low-energy model $\mathcal{H}_{\mathrm{eff}}^{{\rm large} I}$ in the $2^{N/2}$-dimensional subspace without hardcore bosons which can be written purely in terms of the pseudo-spins. Up to second-order perturbation theory in $J/I$ one obtains 
\begin{align}
    \mathcal{H}_{\mathrm{eff}}^{{\rm large} I} = -\frac{IN}{2}-\frac{NJ^2}{16I} -2J \sum_p \hat{\Tilde{B}}_p - \frac{J^2}{4I}\sum_{s} \hat{\Tilde{A}}_s\, .
\end{align}

This effective model corresponds to a single-layer toric code with different couplings in front of star and plaquette operators, which is again exactly solvable. The ground states in the large-Ising phase are characterized by eigenvalues $\Tilde{a}_s=\Tilde{b}_p=1$ $\forall s,p$. Consequently, the ITCB for large $I$ displays $\mathbb{Z}_2$ topological order with the topological entanglement entropy $\gamma_{\rm TC}={\rm log}\,2$ of a single toric code \cite{Castelnovo_2008} and the topologically ordered ground states lie in the parity sector with \mbox{$a_s^= =\Tilde{a}_s=+1$} $\forall s$. Furthermore, higher-order corrections in $J/I$ are always products of effective star and plaquette operators so that the effective low-energy model remains exactly solvable at any order in perturbation theory. This is a consequence of the reduced Hilbert space dimension $2^{N/2}$ of the low-energy subspace, 
    which is similar to the multi-plaquette expansion in the Kitaev's honeycomb model \cite{Schmidt_2008} about the anisotropic limit.

\begin{figure}[t]
        \centering
        \includegraphics[width=0.9\columnwidth]{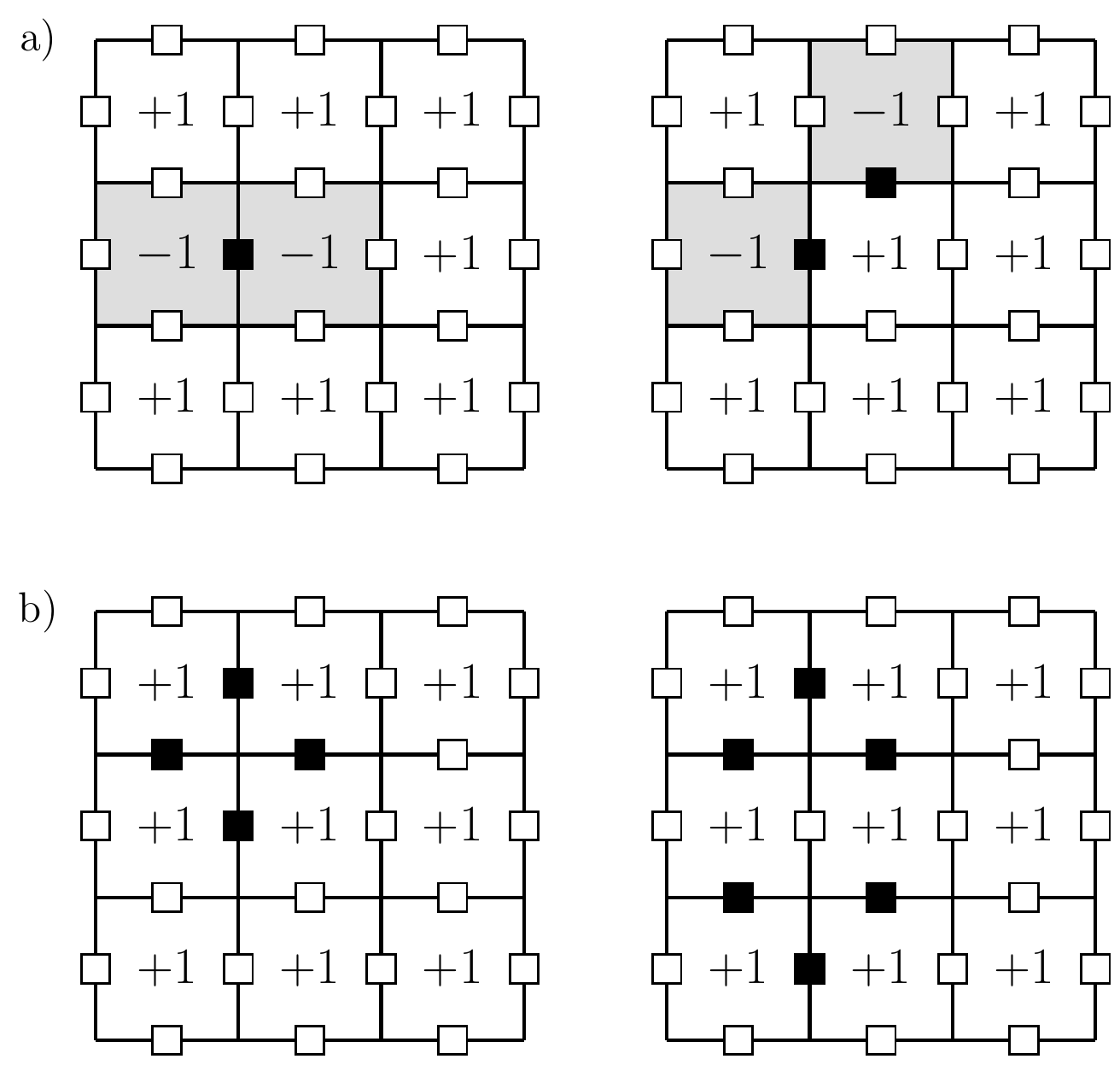}
        \caption{Illustration of the excitations in the large-Ising phase. Filled (empty) squares represent the presence (absence) of a hardcore boson on an Ising dimer. Values $\pm 1$ indicate eigenvalues of $b_{p,2}$. (a) Static one- and two-hardcore boson configurations. (b) Mobile 4- and 6-hardcore boson configurations which are part of the low-energy sector \mbox{$a_s^= = b_{p,\kappa} = 1$} $\forall s,p,\kappa$.}
        \label{Fig::Excitations_Large_Ising}
\end{figure}

Excitations of the low-energy subspace are static given by local charges ($\Tilde{a}_s=-1$) and fluxes ($\Tilde{b}_p=-1$) of the effective toric code. Mobile excitations of the large-Ising phase must therefore involve the hardcore bosons. Interestingly, the mobility of the bosons is strongly constraint by Eq.~\ref{Eq::Bp2} assuming the ground-state sector $\tilde{b}_p=+1$ $\forall p$, since the parity of bosons on each plaquette $p$ is directly linked to the presence or absence of a flux with $b_{p,2}=-1$. Sectors with one, two, or three bosons always involve the presence of fluxes with $b_{p,\kappa}=-1$ (see also Fig.~\ref{Fig::Excitations_Large_Ising}a). Hence, these boson configurations are not allowed to hop. 
    Elementary excitations, which are contained in the low-energy sector with $b_{p,\kappa}=+1$ $\forall p$, are illustrated in Fig.~\ref{Fig::Excitations_Large_Ising}b and consist of four or six hardcore bosons. As we detail below, these excitations are mobile and relevant for the breakdown of the large-Ising phase.

{\it{Low-Ising phase:}} For $I=0$ the system corresponds to two decoupled toric code layers and is exactly solvable, since the $a_{s,\kappa}$ and the $b_{p,\kappa}$ are conserved quantities. Ground states are given by the direct product of the topologically ordered ground states of each toric code layer, which are characterized by eigenvalues $a_{s,\kappa}=b_{p,\kappa}=1$ $\forall s,p,\kappa$
as well as $\mathbb{Z}_2$ eigenvalues of non-local string operators for topologies with genus $g>0$. The number of ground states scales as $4^{2g}$ and the topological entanglement entropy is given by
$\gamma=2\gamma_{\rm TC}$. The ground states lie in the parity sector with \mbox{$a_s^= =  a_{s,1}a_{s,2} =+1$} $\forall s$ and have an energy $E_0=-JN$. Each eigenvalue $-1$ of either $a_{s,\kappa}$ or $b_{p,\kappa}$ costs an energy $2J$. The excitations with $a_{s,\kappa}=-1$ are called charges and the ones with $b_{p,\kappa}=-1$ fluxes. As a consequence, states with an odd number of charges, e.g., single-charge excitations, have at least one parity quantum number $a_s^= =-1$ and are therefore not in the same parity sector as the ground states. The lowest-energy excitation within the ground-state parity sector with \mbox{$a_s^= =+1$} $\forall s$ is a double-charge $a_{s,1}=a_{s,2}=-1$ with excitation energy $4J$.

\begin{figure}[t]
        \centering
        \includegraphics[width=\columnwidth]{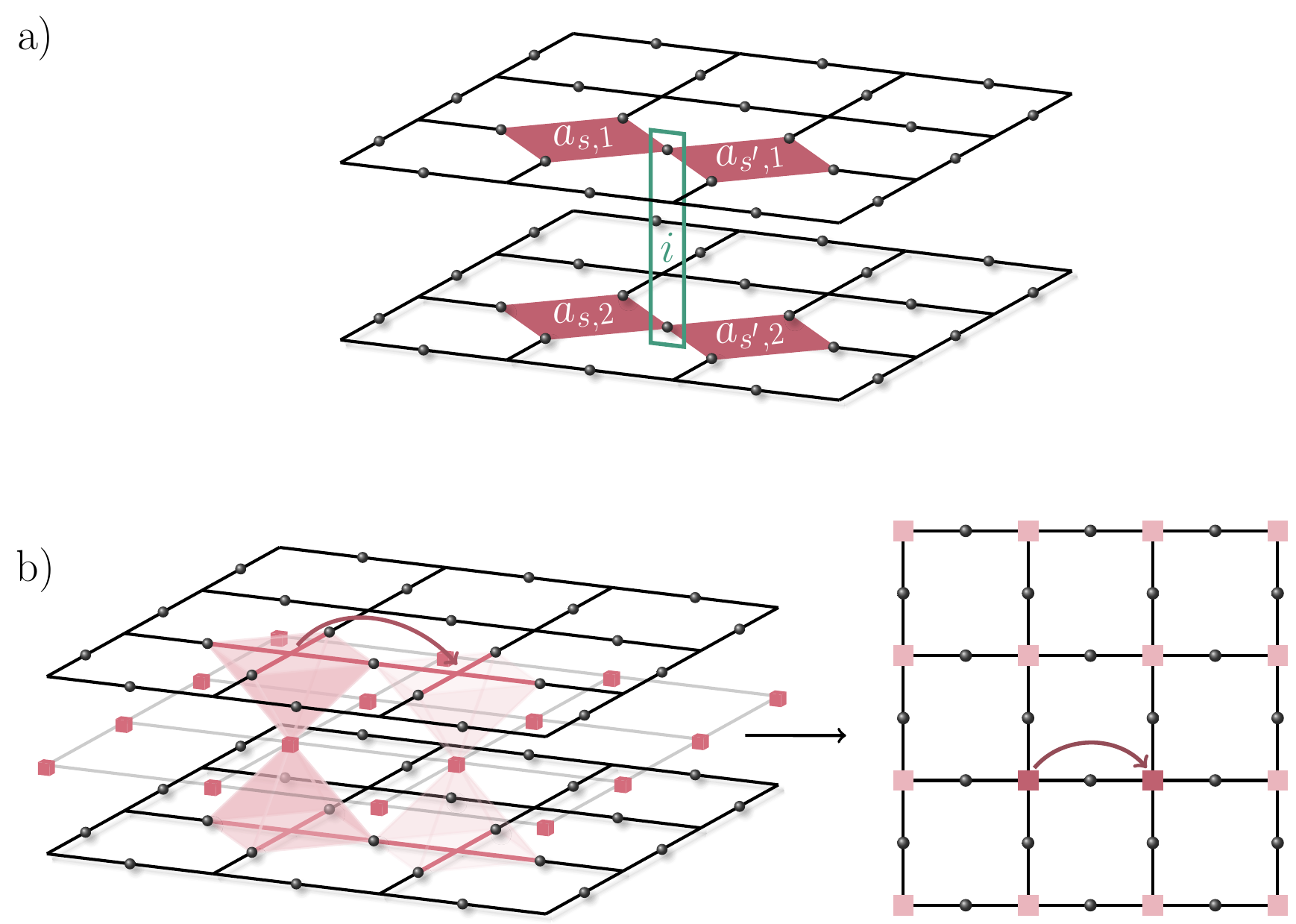}
        \caption{(a) {The Ising interaction on dimer $i$ affects the eigenvalues $a_{s,\kappa}$ and $a_{s',\kappa}$.} (b) Double-charge hopping in the low-Ising phase. On the left this hopping is shown in the original bilayer lattice of the ITCB and on the right as an effective square lattice built by the centers of the double-stars.}
        \label{Fig::Double_charge_hopping}
\end{figure}

For $I>0$, flux excitations with $b_{p,\kappa}=-1$ remain exactly static, since the $b_{p,\kappa}$ are conserved quantities of the ITCB. In contrast, the Ising interaction introduces quantum fluctuations which affect the properties of charge excitations. The Ising interaction $\sigma_{i,1}^z\sigma_{i,2}^z$ on dimer $i$ flips the eigenvalues $a_{s,\kappa}$ of the two pairs of adjacent star operators which contain dimer $i$ (see Fig.~\ref{Fig::Double_charge_hopping}a). Since the local parities $a_s^=$ are conserved, the Ising coupling between the toric code layers changes the number of charges by an even amount locally on each double star $(a_{s,1},a_{s,2})$, i.e.\,, either a double-charge is created or eliminated $(1,1)\leftrightarrow (-1,-1)$ or a charge hops from one layer to the other \mbox{$(1,-1)\leftrightarrow (-1,1)$}. Single charges therefore remain static to any order in perturbation theory, since an effective hopping of a single charge is impossible without changing the conserved parity quantum numbers. In contrast, double-charges are allowed to hop already in first-order perturbation theory in $I/J$ as illustrated in Fig.~\ref{Fig::Double_charge_hopping}b. As we will demonstrate next by an exact duality mapping, it is the condensation of these bosonic double-charges which leads to the breakdown of the $\mathbb{Z}_2\times\mathbb{Z}_2$ intrinsic topological order of the low-Ising phase.

{\it{Duality and phase transition:}} The ground states of the low- and large-Ising phase are contained in the parity sector \mbox{$a_s^= = b_{p,\kappa} = 1$} $\forall s,p,\kappa$. Hence, if the ground-state phase diagram of the ITCB consists only of these two distinct topological phases separated by a single phase transition, the low-energy physics is always contained in this parity sector. All other sectors are indeed not expected to be relevant at low energies, since the presence of some $b_{p,\kappa} = -1$ or $a_s^= = -1$ corresponds to the creation of static excitations, which are not able to reduce their energy significantly by quantum fluctuations.

The only degrees of freedom which fluctuate are the charges due to $b_{p,\kappa} = 1$ $\forall p,\kappa$ and, in this parity sector, the two allowed configurations on a double-star $(a_{s,1},a_{s,2})$ are the absence $(1,1)$ or the presence $(-1,-1)$ of a double-charge. This $\mathbb{Z}_2$ degree of freedom can be represented by a pseudo-spin 1/2 $\vec{\mu}_s$ centered on the star $s$ so that $\ket{\uparrow} \equiv (-1,-1)$ and $\ket{\downarrow} \equiv (1,1)$.

The ITCB in this parity sector can then be mapped to a transverse-field Ising model (TFIM) in terms of the $\vec{\mu}_s$'s on the dual square lattice built by the double-stars
\begin{align}
 \mathcal{H}_{\rm dual}=2J\sum_{s} \mu^{z}_s - I\sum_{\langle s, s^\prime\rangle} \mu^{x}_s\mu^{x}_{s^\prime}-\frac{JN}{2}\, ,
\end{align}
since the sum over the star operators within the ITCB is a counting operator of double-charges, which yields the effective field term $2J\sum_{s} \mu^{z}_s$, and the Ising interaction $\sigma_{i,1}^z\sigma_{i,2}^z$ on dimer $i$ flips the eigenvalues $a_{s,\kappa}$ and $a_{s^\prime,\kappa}$ on two adjacent stars $s,s^\prime$ as discussed before (see Fig.~\ref{Fig::Double_charge_hopping}a) resulting in the effective Ising interaction $\mu^{x}_s\mu^{x}_{s^\prime}$.

The TFIM on the square lattice displays a second-order phase transition at $I/2J=0.3284$ in the 3d Ising universality class separating the high-field polarized phase and the $\mathbb{Z}_2$ symmetry-broken ordered phase \cite{He_1990,Bloete_2002}. Both phases are gapped and the phase transition can be described by the condensation of dressed spin-flip quasiparticles from both sides. The polarized phase of the TFIM is dual to the $\mathbb{Z}_2\times\mathbb{Z}_2$ topologically-ordered low-Ising phase of the ITCB. The energetic properties of the spin-flip quasiparticles of the polarized phase in the TFIM are therefore exactly equivalent to the ones of double-charges in the low-Ising phase. The topological quantum phase transition between the topologically ordered low- and large-Ising phases can be described by the condensation of the bosonic double-charges. Furthermore, the same phase transition can also be understood in terms of condensation when looking at the breakdown of the large-Ising phase. The latter $\mathbb{Z}_2$ topologically-ordered phase is dual to the $\mathbb{Z}_2$ symmetry-broken phase of the TFIM. The elementary excitations of this ordered phase are again dressed spin-flip excitations, which condense at the quantum critical point. The associated elementary excitation of the large-Ising phase in the ITCB correspond to the four-boson excitations on stars as illustrated in Fig.~\ref{Fig::Excitations_Large_Ising}b. Altogether, the breakdown of both topologically-ordered phases of the ITCB can be described by the condensation of bosonic quasi-particles and the quantum phase transition lies in the 3d Ising$^*$ universality class \cite{Schuler_2016}.  

{\it{Topological entanglement entropy:}} The topological entanglement entropy $\gamma$ is a universal correction to the area law in the R\'enyi entropy of order $\alpha$ and signals the presence of intrinsic topological order \cite{Levin_2006,Kitaev_2006_b}. Its value for the $\mathbb{Z}_2$ topologically-ordered phase at values $I/2J>0.3284$ is equal to $\gamma_{\rm TC}$ due to the effective description in terms of a generalized single-layer toric code. For the 
low-Ising phase of the ITCB at values $I/2J<0.3284$, one can use the exact duality to the TFIM to transfer known results for the robustness of the topological entanglement entropy of the single-layer toric code in a $z$-field \cite{Halasz_2012}. Indeed, also the single-layer toric code in a $z$-field can be mapped in the sector without fluxes to a dual TFIM on the square lattice. The essential difference is that in the latter case single charges play the role of double-charges in the ITCB. Consequently, the low-Ising phase is characterized by a topological entanglement entropy of $2\gamma_{\rm TC}$ and corrections to the R\'{e}nyi entropy due to $I$ only give rise to non-universal contributions to the area law. Altogether, the topological entanglement entropy $\gamma$ serves as a non-local order parameter which jumps from $2\gamma_{\rm TC}$ to $\gamma_{\rm TC}$ at the phase transition. 

{\it{Conclusions:}} We have demonstrated that the ITCB represents a paradigmatic system which hosts a second-order phase transition between the two simplest quantum phases with intrinsic topological order. Important physical ingredients to reach this conclusion are the presence of exact local parity quantum numbers, a duality mapping, and the possibility to implement an effective single-layer toric code in the large-Ising limit. The same logic is expected to hold for two copies of any appropriate, exactly solvable, $\mathbb{Z}_2$ stabilizer code like the 3d toric code \cite{Hamma_2005,Nussinov_2008,Reiss_2019} or the X-Cube model \cite{Vijay_2016} coupled by an Ising interaction which commutes with one stabilizer type. However, the resulting nature of the phase transition has to be investigated case by case. Further, the degenerate Ising-dimer limit has also been used by Kitaev to implement the toric code in the more realistic Kitaev honeycomb model \cite{Kitaev_2006}, where interaction frustration of competing Ising interactions gives rise to an effective toric code in fourth-order perturbation theory about the anisotropic Kitaev limit. This suggests that the ITCB can be realized as an effective low-energy model of a bilayer Kitaev honeycomb model, where the layers are coupled in an appropriate fashion. Such a bilayer Kitaev honeycomb model, which is distinct from the Kitaev-Heisenberg bilayer in Ref.~\cite{Seifert_2018}, represents an interesting option for a potential solid-state realization of the ITCB in so-called Kitaev materials.

{\it{Acknowledgments:}}
KPS acknowledges financial support by the German Science Foundation (DFG) through the grant SCHM 2511/11-1.


\begin{thebibliography}{74}%
\makeatletter
\providecommand \@ifxundefined [1]{%
 \@ifx{#1\undefined}
}%
\providecommand \@ifnum [1]{%
 \ifnum #1\expandafter \@firstoftwo
 \else \expandafter \@secondoftwo
 \fi
}%
\providecommand \@ifx [1]{%
 \ifx #1\expandafter \@firstoftwo
 \else \expandafter \@secondoftwo
 \fi
}%
\providecommand \natexlab [1]{#1}%
\providecommand \enquote  [1]{``#1''}%
\providecommand \bibnamefont  [1]{#1}%
\providecommand \bibfnamefont [1]{#1}%
\providecommand \citenamefont [1]{#1}%
\providecommand \href@noop [0]{\@secondoftwo}%
\providecommand \href [0]{\begingroup \@sanitize@url \@href}%
\providecommand \@href[1]{\@@startlink{#1}\@@href}%
\providecommand \@@href[1]{\endgroup#1\@@endlink}%
\providecommand \@sanitize@url [0]{\catcode `\\12\catcode `\$12\catcode
  `\&12\catcode `\#12\catcode `\^12\catcode `\_12\catcode `\%12\relax}%
\providecommand \@@startlink[1]{}%
\providecommand \@@endlink[0]{}%
\providecommand \url  [0]{\begingroup\@sanitize@url \@url }%
\providecommand \@url [1]{\endgroup\@href {#1}{\urlprefix }}%
\providecommand \urlprefix  [0]{URL }%
\providecommand \Eprint [0]{\href }%
\providecommand \doibase [0]{http://dx.doi.org/}%
\providecommand \selectlanguage [0]{\@gobble}%
\providecommand \bibinfo  [0]{\@secondoftwo}%
\providecommand \bibfield  [0]{\@secondoftwo}%
\providecommand \translation [1]{[#1]}%
\providecommand \BibitemOpen [0]{}%
\providecommand \bibitemStop [0]{}%
\providecommand \bibitemNoStop [0]{.\EOS\space}%
\providecommand \EOS [0]{\spacefactor3000\relax}%
\providecommand \BibitemShut  [1]{\csname bibitem#1\endcsname}%
\let\auto@bib@innerbib\@empty
\bibitem [{\citenamefont {Wen}(1989)}]{Wen_1989}%
  \BibitemOpen
  \bibfield  {author} {\bibinfo {author} {\bibfnamefont {X.-G.}\ \bibnamefont
  {Wen}},\ }\href {\doibase 10.1103/PhysRevB.40.7387} {\bibfield  {journal}
  {\bibinfo  {journal} {Phys. Rev. B}\ }\textbf {\bibinfo {volume} {40}},\
  \bibinfo {pages} {7387} (\bibinfo {year} {1989})}\BibitemShut {NoStop}%
\bibitem [{\citenamefont {Wen}(1990)}]{Wen_1990}%
  \BibitemOpen
  \bibfield  {author} {\bibinfo {author} {\bibfnamefont {X.-G.}\ \bibnamefont
  {Wen}},\ }\href {\doibase 10.1142/S0217979290000139} {\bibfield  {journal}
  {\bibinfo  {journal} {Int. J. Mod. Phys. B}\ }\textbf {\bibinfo {volume}
  {4}},\ \bibinfo {pages} {239} (\bibinfo {year} {1990})}\BibitemShut {NoStop}%
\bibitem [{\citenamefont {Wen}(2004)}]{Wen_2004}%
  \BibitemOpen
  \bibfield  {author} {\bibinfo {author} {\bibfnamefont {X.-G.}\ \bibnamefont
  {Wen}},\ }\href@noop {} {\emph {\bibinfo {title} {Quantum Field Theory of
  Many-body Systems: From the Origin of Sound to an Origin of Light and
  Electrons}}}\ (\bibinfo  {publisher} {Oxford University Press},\ \bibinfo
  {year} {2004})\BibitemShut {NoStop}%
\bibitem [{\citenamefont {Leinaas}\ and\ \citenamefont
  {Myrheim}(1977)}]{Leinaas_1977}%
  \BibitemOpen
  \bibfield  {author} {\bibinfo {author} {\bibfnamefont {J.~M.}\ \bibnamefont
  {Leinaas}}\ and\ \bibinfo {author} {\bibfnamefont {J.}~\bibnamefont
  {Myrheim}},\ }\href {\doibase 10.1007/BF02727953} {\bibfield  {journal}
  {\bibinfo  {journal} {Il Nuovo Cimento B}\ }\textbf {\bibinfo {volume}
  {37}},\ \bibinfo {pages} {1} (\bibinfo {year} {1977})}\BibitemShut {NoStop}%
\bibitem [{\citenamefont {Wilczek}(1982)}]{Wilczek_1982}%
  \BibitemOpen
  \bibfield  {author} {\bibinfo {author} {\bibfnamefont {F.}~\bibnamefont
  {Wilczek}},\ }\href {\doibase 10.1103/PhysRevLett.48.1144} {\bibfield
  {journal} {\bibinfo  {journal} {Phys. Rev. Lett.}\ }\textbf {\bibinfo
  {volume} {48}},\ \bibinfo {pages} {1144} (\bibinfo {year}
  {1982})}\BibitemShut {NoStop}%
\bibitem [{\citenamefont {Chamon}(2005)}]{Chamon_2005}%
  \BibitemOpen
  \bibfield  {author} {\bibinfo {author} {\bibfnamefont {C.}~\bibnamefont
  {Chamon}},\ }\href {\doibase 10.1103/PhysRevLett.94.040402} {\bibfield
  {journal} {\bibinfo  {journal} {Phys. Rev. Lett.}\ }\textbf {\bibinfo
  {volume} {94}},\ \bibinfo {pages} {040402} (\bibinfo {year}
  {2005})}\BibitemShut {NoStop}%
\bibitem [{\citenamefont {Bravyi}\ \emph {et~al.}(2011)\citenamefont {Bravyi},
  \citenamefont {Leemhuis},\ and\ \citenamefont {Terhal}}]{Bravyi_2011}%
  \BibitemOpen
  \bibfield  {author} {\bibinfo {author} {\bibfnamefont {S.}~\bibnamefont
  {Bravyi}}, \bibinfo {author} {\bibfnamefont {B.}~\bibnamefont {Leemhuis}}, \
  and\ \bibinfo {author} {\bibfnamefont {B.~M.}\ \bibnamefont {Terhal}},\
  }\href {\doibase 10.1016/j.aop.2010.11.002} {\bibfield  {journal} {\bibinfo
  {journal} {Ann. Phys.}\ }\textbf {\bibinfo {volume} {326}},\ \bibinfo {pages}
  {839} (\bibinfo {year} {2011})}\BibitemShut {NoStop}%
\bibitem [{\citenamefont {Haah}(2011)}]{Haah_2011}%
  \BibitemOpen
  \bibfield  {author} {\bibinfo {author} {\bibfnamefont {J.}~\bibnamefont
  {Haah}},\ }\href {\doibase 10.1103/PhysRevA.83.042330} {\bibfield  {journal}
  {\bibinfo  {journal} {Phys. Rev. A}\ }\textbf {\bibinfo {volume} {83}},\
  \bibinfo {pages} {042330} (\bibinfo {year} {2011})}\BibitemShut {NoStop}%
\bibitem [{\citenamefont {Yoshida}(2013)}]{Yoshida_2013}%
  \BibitemOpen
  \bibfield  {author} {\bibinfo {author} {\bibfnamefont {B.}~\bibnamefont
  {Yoshida}},\ }\href {\doibase 10.1103/PhysRevB.88.125122} {\bibfield
  {journal} {\bibinfo  {journal} {Phys. Rev. B}\ }\textbf {\bibinfo {volume}
  {88}},\ \bibinfo {pages} {125122} (\bibinfo {year} {2013})}\BibitemShut
  {NoStop}%
\bibitem [{\citenamefont {Vijay}\ \emph {et~al.}(2015)\citenamefont {Vijay},
  \citenamefont {Haah},\ and\ \citenamefont {Fu}}]{Vijay_2015}%
  \BibitemOpen
  \bibfield  {author} {\bibinfo {author} {\bibfnamefont {S.}~\bibnamefont
  {Vijay}}, \bibinfo {author} {\bibfnamefont {J.}~\bibnamefont {Haah}}, \ and\
  \bibinfo {author} {\bibfnamefont {L.}~\bibnamefont {Fu}},\ }\href {\doibase
  10.1103/PhysRevB.92.235136} {\bibfield  {journal} {\bibinfo  {journal} {Phys.
  Rev. B}\ }\textbf {\bibinfo {volume} {92}},\ \bibinfo {pages} {235136}
  (\bibinfo {year} {2015})}\BibitemShut {NoStop}%
\bibitem [{\citenamefont {Vijay}\ \emph {et~al.}(2016)\citenamefont {Vijay},
  \citenamefont {Haah},\ and\ \citenamefont {Fu}}]{Vijay_2016}%
  \BibitemOpen
  \bibfield  {author} {\bibinfo {author} {\bibfnamefont {S.}~\bibnamefont
  {Vijay}}, \bibinfo {author} {\bibfnamefont {J.}~\bibnamefont {Haah}}, \ and\
  \bibinfo {author} {\bibfnamefont {L.}~\bibnamefont {Fu}},\ }\href {\doibase
  10.1103/PhysRevB.94.235157} {\bibfield  {journal} {\bibinfo  {journal} {Phys.
  Rev. B}\ }\textbf {\bibinfo {volume} {94}},\ \bibinfo {pages} {235157}
  (\bibinfo {year} {2016})}\BibitemShut {NoStop}%
\bibitem [{\citenamefont {Kitaev}(2003)}]{Kitaev_2003}%
  \BibitemOpen
  \bibfield  {author} {\bibinfo {author} {\bibfnamefont {A.}~\bibnamefont
  {Kitaev}},\ }\href {\doibase 10.1016/S0003-4916(02)00018-0} {\bibfield
  {journal} {\bibinfo  {journal} {Ann. Phys.}\ }\textbf {\bibinfo {volume}
  {303}},\ \bibinfo {pages} {2} (\bibinfo {year} {2003})}\BibitemShut {NoStop}%
\bibitem [{\citenamefont {Nayak}\ \emph {et~al.}(2008)\citenamefont {Nayak},
  \citenamefont {Simon}, \citenamefont {Stern}, \citenamefont {Freedman},\ and\
  \citenamefont {Sarma}}]{Nayak_2008}%
  \BibitemOpen
  \bibfield  {author} {\bibinfo {author} {\bibfnamefont {C.}~\bibnamefont
  {Nayak}}, \bibinfo {author} {\bibfnamefont {S.~H.}\ \bibnamefont {Simon}},
  \bibinfo {author} {\bibfnamefont {A.}~\bibnamefont {Stern}}, \bibinfo
  {author} {\bibfnamefont {M.}~\bibnamefont {Freedman}}, \ and\ \bibinfo
  {author} {\bibfnamefont {S.~D.}\ \bibnamefont {Sarma}},\ }\href {\doibase
  10.1103/RevModPhys.80.1083} {\bibfield  {journal} {\bibinfo  {journal} {Rev.
  Mod. Phys.}\ }\textbf {\bibinfo {volume} {80}},\ \bibinfo {pages} {1083}
  (\bibinfo {year} {2008})}\BibitemShut {NoStop}%
\bibitem [{\citenamefont {Laughlin}(1983)}]{Laughlin_1983}%
  \BibitemOpen
  \bibfield  {author} {\bibinfo {author} {\bibfnamefont {R.~B.}\ \bibnamefont
  {Laughlin}},\ }\href {\doibase 10.1103/PhysRevLett.50.1395} {\bibfield
  {journal} {\bibinfo  {journal} {Phys. Rev. Lett.}\ }\textbf {\bibinfo
  {volume} {50}},\ \bibinfo {pages} {1395} (\bibinfo {year}
  {1983})}\BibitemShut {NoStop}%
\bibitem [{\citenamefont {Tsui}\ \emph {et~al.}(1982)\citenamefont {Tsui},
  \citenamefont {Stormer},\ and\ \citenamefont {Gossard}}]{Tsui_1982}%
  \BibitemOpen
  \bibfield  {author} {\bibinfo {author} {\bibfnamefont {D.}~\bibnamefont
  {Tsui}}, \bibinfo {author} {\bibfnamefont {H.}~\bibnamefont {Stormer}}, \
  and\ \bibinfo {author} {\bibfnamefont {A.}~\bibnamefont {Gossard}},\ }\href
  {\doibase 10.1103/PhysRevLett.48.1559} {\bibfield  {journal} {\bibinfo
  {journal} {Phys. Rev. Lett.}\ }\textbf {\bibinfo {volume} {48}},\ \bibinfo
  {pages} {1559} (\bibinfo {year} {1982})}\BibitemShut {NoStop}%
\bibitem [{\citenamefont {Balents}(2010)}]{Balents_2010}%
  \BibitemOpen
  \bibfield  {author} {\bibinfo {author} {\bibfnamefont {L.}~\bibnamefont
  {Balents}},\ }\href {\doibase 10.1038/nature08917} {\bibfield  {journal}
  {\bibinfo  {journal} {Nature}\ }\textbf {\bibinfo {volume} {464}},\ \bibinfo
  {pages} {199} (\bibinfo {year} {2010})}\BibitemShut {NoStop}%
\bibitem [{\citenamefont {Chaloupka}\ \emph {et~al.}(2010)\citenamefont
  {Chaloupka}, \citenamefont {Jackeli},\ and\ \citenamefont
  {Khaliullin}}]{Jackeli_2010}%
  \BibitemOpen
  \bibfield  {author} {\bibinfo {author} {\bibfnamefont {J.}~\bibnamefont
  {Chaloupka}}, \bibinfo {author} {\bibfnamefont {G.}~\bibnamefont {Jackeli}},
  \ and\ \bibinfo {author} {\bibfnamefont {G.}~\bibnamefont {Khaliullin}},\
  }\href {\doibase 10.1103/PhysRevLett.105.027204} {\bibfield  {journal}
  {\bibinfo  {journal} {Phys. Rev. Lett.}\ }\textbf {\bibinfo {volume} {105}},\
  \bibinfo {pages} {027204} (\bibinfo {year} {2010})}\BibitemShut {NoStop}%
\bibitem [{\citenamefont {Singh}\ \emph {et~al.}(2012)\citenamefont {Singh},
  \citenamefont {Manni}, \citenamefont {Reuther}, \citenamefont {Berlijn},
  \citenamefont {Thomale}, \citenamefont {Ku}, \citenamefont {Trebst},\ and\
  \citenamefont {Gegenwart}}]{Singh_2012}%
  \BibitemOpen
  \bibfield  {author} {\bibinfo {author} {\bibfnamefont {Y.}~\bibnamefont
  {Singh}}, \bibinfo {author} {\bibfnamefont {S.}~\bibnamefont {Manni}},
  \bibinfo {author} {\bibfnamefont {J.}~\bibnamefont {Reuther}}, \bibinfo
  {author} {\bibfnamefont {T.}~\bibnamefont {Berlijn}}, \bibinfo {author}
  {\bibfnamefont {R.}~\bibnamefont {Thomale}}, \bibinfo {author} {\bibfnamefont
  {W.}~\bibnamefont {Ku}}, \bibinfo {author} {\bibfnamefont {S.}~\bibnamefont
  {Trebst}}, \ and\ \bibinfo {author} {\bibfnamefont {P.}~\bibnamefont
  {Gegenwart}},\ }\href {\doibase 10.1103/PhysRevLett.108.127203} {\bibfield
  {journal} {\bibinfo  {journal} {Phys. Rev. Lett.}\ }\textbf {\bibinfo
  {volume} {108}},\ \bibinfo {pages} {127203} (\bibinfo {year}
  {2012})}\BibitemShut {NoStop}%
\bibitem [{\citenamefont {Plumb}\ \emph {et~al.}(2014)\citenamefont {Plumb},
  \citenamefont {Clancy}, \citenamefont {Sandilands}, \citenamefont {Shankar},
  \citenamefont {Hu}, \citenamefont {Burch}, \citenamefont {Kee},\ and\
  \citenamefont {Kim}}]{Plumb_2014}%
  \BibitemOpen
  \bibfield  {author} {\bibinfo {author} {\bibfnamefont {K.~W.}\ \bibnamefont
  {Plumb}}, \bibinfo {author} {\bibfnamefont {J.~P.}\ \bibnamefont {Clancy}},
  \bibinfo {author} {\bibfnamefont {L.~J.}\ \bibnamefont {Sandilands}},
  \bibinfo {author} {\bibfnamefont {V.~V.}\ \bibnamefont {Shankar}}, \bibinfo
  {author} {\bibfnamefont {Y.~F.}\ \bibnamefont {Hu}}, \bibinfo {author}
  {\bibfnamefont {K.~S.}\ \bibnamefont {Burch}}, \bibinfo {author}
  {\bibfnamefont {H.-Y.}\ \bibnamefont {Kee}}, \ and\ \bibinfo {author}
  {\bibfnamefont {Y.-J.}\ \bibnamefont {Kim}},\ }\href {\doibase
  10.1103/PhysRevB.90.041112} {\bibfield  {journal} {\bibinfo  {journal} {Phys.
  Rev. B}\ }\textbf {\bibinfo {volume} {90}},\ \bibinfo {pages} {041112}
  (\bibinfo {year} {2014})}\BibitemShut {NoStop}%
\bibitem [{\citenamefont {Banerjee}\ \emph {et~al.}(2017)\citenamefont
  {Banerjee}, \citenamefont {Yan}, \citenamefont {Knolle}, \citenamefont
  {Bridges}, \citenamefont {Stone}, \citenamefont {Lumsden}, \citenamefont
  {Mandrus}, \citenamefont {Tennant}, \citenamefont {Moessner},\ and\
  \citenamefont {Nagler1}}]{Banerjee_2017}%
  \BibitemOpen
  \bibfield  {author} {\bibinfo {author} {\bibfnamefont {A.}~\bibnamefont
  {Banerjee}}, \bibinfo {author} {\bibfnamefont {J.}~\bibnamefont {Yan}},
  \bibinfo {author} {\bibfnamefont {J.}~\bibnamefont {Knolle}}, \bibinfo
  {author} {\bibfnamefont {C.~A.}\ \bibnamefont {Bridges}}, \bibinfo {author}
  {\bibfnamefont {M.~B.}\ \bibnamefont {Stone}}, \bibinfo {author}
  {\bibfnamefont {M.}~\bibnamefont {Lumsden}}, \bibinfo {author} {\bibfnamefont
  {D.~G.}\ \bibnamefont {Mandrus}}, \bibinfo {author} {\bibfnamefont
  {D.}~\bibnamefont {Tennant}}, \bibinfo {author} {\bibfnamefont
  {R.}~\bibnamefont {Moessner}}, \ and\ \bibinfo {author} {\bibfnamefont
  {S.~E.}\ \bibnamefont {Nagler1}},\ }\href {\doibase 10.1126/science.aah6015}
  {\bibfield  {journal} {\bibinfo  {journal} {Science}\ }\textbf {\bibinfo
  {volume} {356}},\ \bibinfo {pages} {1055} (\bibinfo {year}
  {2017})}\BibitemShut {NoStop}%
\bibitem [{\citenamefont {Savary}\ and\ \citenamefont
  {Balents}(2017)}]{Savary_2017}%
  \BibitemOpen
  \bibfield  {author} {\bibinfo {author} {\bibfnamefont {L.}~\bibnamefont
  {Savary}}\ and\ \bibinfo {author} {\bibfnamefont {L.}~\bibnamefont
  {Balents}},\ }\href {\doibase 10.1088/0034-4885/80/1/016502} {\bibfield
  {journal} {\bibinfo  {journal} {Rep. Prog. Phys.}\ }\textbf {\bibinfo
  {volume} {80}},\ \bibinfo {pages} {016502} (\bibinfo {year}
  {2017})}\BibitemShut {NoStop}%
\bibitem [{\citenamefont {Banerjee}\ \emph {et~al.}(2018)\citenamefont
  {Banerjee}, \citenamefont {Lampen-Kelley}, \citenamefont {Knolle},
  \citenamefont {Balz}, \citenamefont {Aczel}, \citenamefont {Winn},
  \citenamefont {Liu}, \citenamefont {Pajerowski}, \citenamefont {Yan},
  \citenamefont {Bridges}, \citenamefont {Savici}, \citenamefont {Chakoumakos},
  \citenamefont {Lumsden}, \citenamefont {Tennant}, \citenamefont {Moessner},
  \citenamefont {Mandrus},\ and\ \citenamefont {Nagler}}]{Banerjee_2018}%
  \BibitemOpen
  \bibfield  {author} {\bibinfo {author} {\bibfnamefont {A.}~\bibnamefont
  {Banerjee}}, \bibinfo {author} {\bibfnamefont {P.}~\bibnamefont
  {Lampen-Kelley}}, \bibinfo {author} {\bibfnamefont {J.}~\bibnamefont
  {Knolle}}, \bibinfo {author} {\bibfnamefont {C.}~\bibnamefont {Balz}},
  \bibinfo {author} {\bibfnamefont {A.~A.}\ \bibnamefont {Aczel}}, \bibinfo
  {author} {\bibfnamefont {B.}~\bibnamefont {Winn}}, \bibinfo {author}
  {\bibfnamefont {Y.}~\bibnamefont {Liu}}, \bibinfo {author} {\bibfnamefont
  {D.}~\bibnamefont {Pajerowski}}, \bibinfo {author} {\bibfnamefont
  {J.}~\bibnamefont {Yan}}, \bibinfo {author} {\bibfnamefont {C.~A.}\
  \bibnamefont {Bridges}}, \bibinfo {author} {\bibfnamefont {A.~T.}\
  \bibnamefont {Savici}}, \bibinfo {author} {\bibfnamefont {B.~C.}\
  \bibnamefont {Chakoumakos}}, \bibinfo {author} {\bibfnamefont {M.~D.}\
  \bibnamefont {Lumsden}}, \bibinfo {author} {\bibfnamefont {D.~A.}\
  \bibnamefont {Tennant}}, \bibinfo {author} {\bibfnamefont {R.}~\bibnamefont
  {Moessner}}, \bibinfo {author} {\bibfnamefont {D.~G.}\ \bibnamefont
  {Mandrus}}, \ and\ \bibinfo {author} {\bibfnamefont {S.~E.}\ \bibnamefont
  {Nagler}},\ }\href {\doibase 10.1038/s41535-018-0079-2} {\bibfield  {journal}
  {\bibinfo  {journal} {NPJ Quantum Materials}\ }\textbf {\bibinfo {volume}
  {3}} (\bibinfo {year} {2018}),\ 10.1038/s41535-018-0079-2}\BibitemShut
  {NoStop}%
\bibitem [{\citenamefont {Han}\ \emph {et~al.}(2007)\citenamefont {Han},
  \citenamefont {Raussendorf},\ and\ \citenamefont {Duan}}]{Han_2007}%
  \BibitemOpen
  \bibfield  {author} {\bibinfo {author} {\bibfnamefont {Y.-J.}\ \bibnamefont
  {Han}}, \bibinfo {author} {\bibfnamefont {R.}~\bibnamefont {Raussendorf}}, \
  and\ \bibinfo {author} {\bibfnamefont {L.~M.}\ \bibnamefont {Duan}},\ }\href
  {\doibase 10.1103/PhysRevLett.98.150404} {\bibfield  {journal} {\bibinfo
  {journal} {Phys. Rev. Lett.}\ }\textbf {\bibinfo {volume} {98}},\ \bibinfo
  {pages} {150404} (\bibinfo {year} {2007})}\BibitemShut {NoStop}%
\bibitem [{\citenamefont {Lu}\ \emph {et~al.}(2009)\citenamefont {Lu},
  \citenamefont {Gao}, \citenamefont {G\"{u}hne}, \citenamefont {Zhou},
  \citenamefont {Chen},\ and\ \citenamefont {Pan}}]{Lu_2009}%
  \BibitemOpen
  \bibfield  {author} {\bibinfo {author} {\bibfnamefont {C.-Y.}\ \bibnamefont
  {Lu}}, \bibinfo {author} {\bibfnamefont {W.-B.}\ \bibnamefont {Gao}},
  \bibinfo {author} {\bibfnamefont {O.}~\bibnamefont {G\"{u}hne}}, \bibinfo
  {author} {\bibfnamefont {X.-Q.}\ \bibnamefont {Zhou}}, \bibinfo {author}
  {\bibfnamefont {Z.-B.}\ \bibnamefont {Chen}}, \ and\ \bibinfo {author}
  {\bibfnamefont {J.-W.}\ \bibnamefont {Pan}},\ }\href {\doibase
  10.1103/PhysRevLett.102.030502} {\bibfield  {journal} {\bibinfo  {journal}
  {Phys. Rev. Lett.}\ }\textbf {\bibinfo {volume} {102}},\ \bibinfo {pages}
  {030502} (\bibinfo {year} {2009})}\BibitemShut {NoStop}%
\bibitem [{\citenamefont {Pachos}\ \emph {et~al.}(2009)\citenamefont {Pachos},
  \citenamefont {Wieczorek}, \citenamefont {Schmid}, \citenamefont {Kiesel},
  \citenamefont {Pohlner},\ and\ \citenamefont {Weinfurter}}]{Pachos_2009}%
  \BibitemOpen
  \bibfield  {author} {\bibinfo {author} {\bibfnamefont {J.~K.}\ \bibnamefont
  {Pachos}}, \bibinfo {author} {\bibfnamefont {W.}~\bibnamefont {Wieczorek}},
  \bibinfo {author} {\bibfnamefont {C.}~\bibnamefont {Schmid}}, \bibinfo
  {author} {\bibfnamefont {N.}~\bibnamefont {Kiesel}}, \bibinfo {author}
  {\bibfnamefont {R.}~\bibnamefont {Pohlner}}, \ and\ \bibinfo {author}
  {\bibfnamefont {H.}~\bibnamefont {Weinfurter}},\ }\href {\doibase
  10.1088/1367-2630/11/8/083010} {\bibfield  {journal} {\bibinfo  {journal}
  {New J. Phys.}\ }\textbf {\bibinfo {volume} {11}},\ \bibinfo {pages} {083010}
  (\bibinfo {year} {2009})}\BibitemShut {NoStop}%
\bibitem [{\citenamefont {Du}\ \emph {et~al.}(2007)\citenamefont {Du},
  \citenamefont {Zhu}, \citenamefont {Hu},\ and\ \citenamefont
  {Chen}}]{Du_2007}%
  \BibitemOpen
  \bibfield  {author} {\bibinfo {author} {\bibfnamefont {J.-F.}\ \bibnamefont
  {Du}}, \bibinfo {author} {\bibfnamefont {J.}~\bibnamefont {Zhu}}, \bibinfo
  {author} {\bibfnamefont {M.-G.}\ \bibnamefont {Hu}}, \ and\ \bibinfo {author}
  {\bibfnamefont {J.-L.}\ \bibnamefont {Chen}},\ }\href@noop {} {\bibfield
  {journal} {\bibinfo  {journal} {ar{X}iv:0712.2694}\ } (\bibinfo {year}
  {2007})}\BibitemShut {NoStop}%
\bibitem [{\citenamefont {Feng}\ \emph {et~al.}(2013)\citenamefont {Feng},
  \citenamefont {Long},\ and\ \citenamefont {Laflamme}}]{Feng_2012}%
  \BibitemOpen
  \bibfield  {author} {\bibinfo {author} {\bibfnamefont {G.}~\bibnamefont
  {Feng}}, \bibinfo {author} {\bibfnamefont {G.}~\bibnamefont {Long}}, \ and\
  \bibinfo {author} {\bibfnamefont {R.}~\bibnamefont {Laflamme}},\ }\href
  {\doibase 10.1103/PhysRevA.88.022305} {\bibfield  {journal} {\bibinfo
  {journal} {Phys. Rev. A}\ }\textbf {\bibinfo {volume} {88}},\ \bibinfo
  {pages} {022305} (\bibinfo {year} {2013})}\BibitemShut {NoStop}%
\bibitem [{\citenamefont {Peng}\ \emph {et~al.}(2014)\citenamefont {Peng},
  \citenamefont {Luo}, \citenamefont {Zheng}, \citenamefont {Kou},
  \citenamefont {Suter},\ and\ \citenamefont {Du}}]{Peng_2014}%
  \BibitemOpen
  \bibfield  {author} {\bibinfo {author} {\bibfnamefont {X.}~\bibnamefont
  {Peng}}, \bibinfo {author} {\bibfnamefont {Z.}~\bibnamefont {Luo}}, \bibinfo
  {author} {\bibfnamefont {W.}~\bibnamefont {Zheng}}, \bibinfo {author}
  {\bibfnamefont {S.}~\bibnamefont {Kou}}, \bibinfo {author} {\bibfnamefont
  {D.}~\bibnamefont {Suter}}, \ and\ \bibinfo {author} {\bibfnamefont
  {J.}~\bibnamefont {Du}},\ }\href {\doibase 10.1103/PhysRevLett.113.080404}
  {\bibfield  {journal} {\bibinfo  {journal} {Phys. Rev. Lett.}\ }\textbf
  {\bibinfo {volume} {113}},\ \bibinfo {pages} {080404} (\bibinfo {year}
  {2014})}\BibitemShut {NoStop}%
\bibitem [{\citenamefont {Micheli}\ \emph {et~al.}(2006)\citenamefont
  {Micheli}, \citenamefont {Brennen},\ and\ \citenamefont
  {Zoller}}]{Micheli_2006}%
  \BibitemOpen
  \bibfield  {author} {\bibinfo {author} {\bibfnamefont {A.}~\bibnamefont
  {Micheli}}, \bibinfo {author} {\bibfnamefont {G.~K.}\ \bibnamefont
  {Brennen}}, \ and\ \bibinfo {author} {\bibfnamefont {P.}~\bibnamefont
  {Zoller}},\ }\href {\doibase 10.1038/nphys287} {\bibfield  {journal}
  {\bibinfo  {journal} {Nature Phys.}\ }\textbf {\bibinfo {volume} {2}},\
  \bibinfo {pages} {341} (\bibinfo {year} {2006})}\BibitemShut {NoStop}%
\bibitem [{\citenamefont {Paredes}\ and\ \citenamefont
  {Bloch}(2008)}]{Paredes_2008}%
  \BibitemOpen
  \bibfield  {author} {\bibinfo {author} {\bibfnamefont {B.}~\bibnamefont
  {Paredes}}\ and\ \bibinfo {author} {\bibfnamefont {I.}~\bibnamefont
  {Bloch}},\ }\href {\doibase 10.1103/PhysRevA.77.023603} {\bibfield  {journal}
  {\bibinfo  {journal} {Phys. Rev. A}\ }\textbf {\bibinfo {volume} {77}},\
  \bibinfo {pages} {023603} (\bibinfo {year} {2008})}\BibitemShut {NoStop}%
\bibitem [{\citenamefont {Sameti}\ \emph {et~al.}(2017)\citenamefont {Sameti},
  \citenamefont {Poto\ifmmode~\check{c}\else \v{c}\fi{}nik}, \citenamefont
  {Browne}, \citenamefont {Wallraff},\ and\ \citenamefont
  {Hartmann}}]{Sameti_2017}%
  \BibitemOpen
  \bibfield  {author} {\bibinfo {author} {\bibfnamefont {M.}~\bibnamefont
  {Sameti}}, \bibinfo {author} {\bibfnamefont {A.}~\bibnamefont
  {Poto\ifmmode~\check{c}\else \v{c}\fi{}nik}}, \bibinfo {author}
  {\bibfnamefont {D.~E.}\ \bibnamefont {Browne}}, \bibinfo {author}
  {\bibfnamefont {A.}~\bibnamefont {Wallraff}}, \ and\ \bibinfo {author}
  {\bibfnamefont {M.~J.}\ \bibnamefont {Hartmann}},\ }\href {\doibase
  10.1103/PhysRevA.95.042330} {\bibfield  {journal} {\bibinfo  {journal} {Phys.
  Rev. A}\ }\textbf {\bibinfo {volume} {95}},\ \bibinfo {pages} {042330}
  (\bibinfo {year} {2017})}\BibitemShut {NoStop}%
\bibitem [{\citenamefont {Bais}\ \emph {et~al.}(2002)\citenamefont {Bais},
  \citenamefont {Schroers},\ and\ \citenamefont {Slingerland}}]{Bais_2002}%
  \BibitemOpen
  \bibfield  {author} {\bibinfo {author} {\bibfnamefont {F.~A.}\ \bibnamefont
  {Bais}}, \bibinfo {author} {\bibfnamefont {B.~J.}\ \bibnamefont {Schroers}},
  \ and\ \bibinfo {author} {\bibfnamefont {J.~K.}\ \bibnamefont
  {Slingerland}},\ }\href {\doibase 10.1103/PhysRevLett.89.181601} {\bibfield
  {journal} {\bibinfo  {journal} {Phys. Rev. Lett.}\ }\textbf {\bibinfo
  {volume} {89}},\ \bibinfo {pages} {181601} (\bibinfo {year}
  {2002})}\BibitemShut {NoStop}%
\bibitem [{\citenamefont {Bais}\ and\ \citenamefont {Mathy}(2007)}]{Bais_2007}%
  \BibitemOpen
  \bibfield  {author} {\bibinfo {author} {\bibfnamefont {F.}~\bibnamefont
  {Bais}}\ and\ \bibinfo {author} {\bibfnamefont {C.}~\bibnamefont {Mathy}},\
  }\href {\doibase https://doi.org/10.1016/j.aop.2006.05.010} {\bibfield
  {journal} {\bibinfo  {journal} {Annals of Physics}\ }\textbf {\bibinfo
  {volume} {322}},\ \bibinfo {pages} {552 } (\bibinfo {year}
  {2007})}\BibitemShut {NoStop}%
\bibitem [{\citenamefont {Bais}\ and\ \citenamefont
  {Slingerland}(2009)}]{Bais_2009}%
  \BibitemOpen
  \bibfield  {author} {\bibinfo {author} {\bibfnamefont {F.~A.}\ \bibnamefont
  {Bais}}\ and\ \bibinfo {author} {\bibfnamefont {J.~K.}\ \bibnamefont
  {Slingerland}},\ }\href {\doibase 10.1103/PhysRevB.79.045316} {\bibfield
  {journal} {\bibinfo  {journal} {Phys. Rev. B}\ }\textbf {\bibinfo {volume}
  {79}},\ \bibinfo {pages} {045316} (\bibinfo {year} {2009})}\BibitemShut
  {NoStop}%
\bibitem [{\citenamefont {Burnell}\ \emph {et~al.}(2011)\citenamefont
  {Burnell}, \citenamefont {Simon},\ and\ \citenamefont
  {Slingerland}}]{Burnell_2011}%
  \BibitemOpen
  \bibfield  {author} {\bibinfo {author} {\bibfnamefont {F.~J.}\ \bibnamefont
  {Burnell}}, \bibinfo {author} {\bibfnamefont {S.~H.}\ \bibnamefont {Simon}},
  \ and\ \bibinfo {author} {\bibfnamefont {J.~K.}\ \bibnamefont
  {Slingerland}},\ }\href {\doibase 10.1103/PhysRevB.84.125434} {\bibfield
  {journal} {\bibinfo  {journal} {Phys. Rev. B}\ }\textbf {\bibinfo {volume}
  {84}},\ \bibinfo {pages} {125434} (\bibinfo {year} {2011})}\BibitemShut
  {NoStop}%
\bibitem [{\citenamefont {Burnell}(2018)}]{Burnell_2018}%
  \BibitemOpen
  \bibfield  {author} {\bibinfo {author} {\bibfnamefont {F.}~\bibnamefont
  {Burnell}},\ }\href {\doibase 10.1146/annurev-conmatphys-033117-054154}
  {\bibfield  {journal} {\bibinfo  {journal} {Annu. Rev. Condens. Matter
  Phys.}\ }\textbf {\bibinfo {volume} {9}},\ \bibinfo {pages} {307} (\bibinfo
  {year} {2018})}\BibitemShut {NoStop}%
\bibitem [{\citenamefont {Trebst}\ \emph {et~al.}(2007)\citenamefont {Trebst},
  \citenamefont {Werner}, \citenamefont {Troyer}, \citenamefont {Shtengel},\
  and\ \citenamefont {Nayak}}]{Trebst_2007}%
  \BibitemOpen
  \bibfield  {author} {\bibinfo {author} {\bibfnamefont {S.}~\bibnamefont
  {Trebst}}, \bibinfo {author} {\bibfnamefont {P.}~\bibnamefont {Werner}},
  \bibinfo {author} {\bibfnamefont {M.}~\bibnamefont {Troyer}}, \bibinfo
  {author} {\bibfnamefont {K.}~\bibnamefont {Shtengel}}, \ and\ \bibinfo
  {author} {\bibfnamefont {C.}~\bibnamefont {Nayak}},\ }\href {\doibase
  10.1103/PhysRevLett.98.070602} {\bibfield  {journal} {\bibinfo  {journal}
  {Phys. Rev. Lett.}\ }\textbf {\bibinfo {volume} {98}},\ \bibinfo {pages}
  {070602} (\bibinfo {year} {2007})}\BibitemShut {NoStop}%
\bibitem [{\citenamefont {Hamma}\ and\ \citenamefont
  {Lidar}(2008)}]{Hamma_2008_b}%
  \BibitemOpen
  \bibfield  {author} {\bibinfo {author} {\bibfnamefont {A.}~\bibnamefont
  {Hamma}}\ and\ \bibinfo {author} {\bibfnamefont {D.~A.}\ \bibnamefont
  {Lidar}},\ }\href {\doibase 10.1103/PhysRevLett.100.030502} {\bibfield
  {journal} {\bibinfo  {journal} {Phys. Rev. Lett.}\ }\textbf {\bibinfo
  {volume} {100}},\ \bibinfo {pages} {030502} (\bibinfo {year}
  {2008})}\BibitemShut {NoStop}%
\bibitem [{\citenamefont {Yu}\ \emph {et~al.}(2008)\citenamefont {Yu},
  \citenamefont {Kou},\ and\ \citenamefont {Wen}}]{Yu_2008}%
  \BibitemOpen
  \bibfield  {author} {\bibinfo {author} {\bibfnamefont {J.}~\bibnamefont
  {Yu}}, \bibinfo {author} {\bibfnamefont {S.-P.}\ \bibnamefont {Kou}}, \ and\
  \bibinfo {author} {\bibfnamefont {X.-G.}\ \bibnamefont {Wen}},\ }\href
  {\doibase 10.1209/0295-5075/84/17004} {\bibfield  {journal} {\bibinfo
  {journal} {Eur. Phys. Lett.}\ }\textbf {\bibinfo {volume} {84}},\ \bibinfo
  {pages} {17004} (\bibinfo {year} {2008})}\BibitemShut {NoStop}%
\bibitem [{\citenamefont {Vidal}\ \emph
  {et~al.}(2009{\natexlab{a}})\citenamefont {Vidal}, \citenamefont {Dusuel},\
  and\ \citenamefont {Schmidt}}]{Vidal_2009}%
  \BibitemOpen
  \bibfield  {author} {\bibinfo {author} {\bibfnamefont {J.}~\bibnamefont
  {Vidal}}, \bibinfo {author} {\bibfnamefont {S.}~\bibnamefont {Dusuel}}, \
  and\ \bibinfo {author} {\bibfnamefont {K.~P.}\ \bibnamefont {Schmidt}},\
  }\href {\doibase 10.1103/PhysRevB.79.033109} {\bibfield  {journal} {\bibinfo
  {journal} {Phys. Rev. B}\ }\textbf {\bibinfo {volume} {79}},\ \bibinfo
  {pages} {033109} (\bibinfo {year} {2009}{\natexlab{a}})}\BibitemShut
  {NoStop}%
\bibitem [{\citenamefont {Vidal}\ \emph
  {et~al.}(2009{\natexlab{b}})\citenamefont {Vidal}, \citenamefont {Thomale},
  \citenamefont {Schmidt},\ and\ \citenamefont {Dusuel}}]{Vidal_2011}%
  \BibitemOpen
  \bibfield  {author} {\bibinfo {author} {\bibfnamefont {J.}~\bibnamefont
  {Vidal}}, \bibinfo {author} {\bibfnamefont {R.}~\bibnamefont {Thomale}},
  \bibinfo {author} {\bibfnamefont {K.~P.}\ \bibnamefont {Schmidt}}, \ and\
  \bibinfo {author} {\bibfnamefont {S.}~\bibnamefont {Dusuel}},\ }\href
  {\doibase 10.1103/PhysRevB.80.081104} {\bibfield  {journal} {\bibinfo
  {journal} {Phys. Rev. B}\ }\textbf {\bibinfo {volume} {80}},\ \bibinfo
  {pages} {081104} (\bibinfo {year} {2009}{\natexlab{b}})}\BibitemShut
  {NoStop}%
\bibitem [{\citenamefont {Dusuel}\ \emph {et~al.}(2010)\citenamefont {Dusuel},
  \citenamefont {Kamfor}, \citenamefont {Schmidt}, \citenamefont {Thomale},\
  and\ \citenamefont {Vidal}}]{Dusuel_2009}%
  \BibitemOpen
  \bibfield  {author} {\bibinfo {author} {\bibfnamefont {S.}~\bibnamefont
  {Dusuel}}, \bibinfo {author} {\bibfnamefont {M.}~\bibnamefont {Kamfor}},
  \bibinfo {author} {\bibfnamefont {K.~P.}\ \bibnamefont {Schmidt}}, \bibinfo
  {author} {\bibfnamefont {R.}~\bibnamefont {Thomale}}, \ and\ \bibinfo
  {author} {\bibfnamefont {J.}~\bibnamefont {Vidal}},\ }\href {\doibase
  10.1103/PhysRevB.81.064412} {\bibfield  {journal} {\bibinfo  {journal} {Phys.
  Rev. B}\ }\textbf {\bibinfo {volume} {81}},\ \bibinfo {pages} {064412}
  (\bibinfo {year} {2010})}\BibitemShut {NoStop}%
\bibitem [{\citenamefont {Tupitsyn}\ \emph {et~al.}(2010)\citenamefont
  {Tupitsyn}, \citenamefont {Kitaev}, \citenamefont {Prokof'ev},\ and\
  \citenamefont {Stamp}}]{Tupitsyn_2010}%
  \BibitemOpen
  \bibfield  {author} {\bibinfo {author} {\bibfnamefont {I.~S.}\ \bibnamefont
  {Tupitsyn}}, \bibinfo {author} {\bibfnamefont {A.}~\bibnamefont {Kitaev}},
  \bibinfo {author} {\bibfnamefont {N.~V.}\ \bibnamefont {Prokof'ev}}, \ and\
  \bibinfo {author} {\bibfnamefont {P.~C.~E.}\ \bibnamefont {Stamp}},\ }\href
  {\doibase 10.1103/PhysRevB.82.085114} {\bibfield  {journal} {\bibinfo
  {journal} {Phys. Rev. B}\ }\textbf {\bibinfo {volume} {82}},\ \bibinfo
  {pages} {085114} (\bibinfo {year} {2010})}\BibitemShut {NoStop}%
\bibitem [{\citenamefont {Wu}\ \emph {et~al.}(2012)\citenamefont {Wu},
  \citenamefont {Deng},\ and\ \citenamefont {Prokof'ev}}]{Wu_2012}%
  \BibitemOpen
  \bibfield  {author} {\bibinfo {author} {\bibfnamefont {F.}~\bibnamefont
  {Wu}}, \bibinfo {author} {\bibfnamefont {Y.}~\bibnamefont {Deng}}, \ and\
  \bibinfo {author} {\bibfnamefont {N.}~\bibnamefont {Prokof'ev}},\ }\href
  {\doibase 10.1103/PhysRevB.85.195104} {\bibfield  {journal} {\bibinfo
  {journal} {Phys. Rev. B}\ }\textbf {\bibinfo {volume} {85}},\ \bibinfo
  {pages} {195104} (\bibinfo {year} {2012})}\BibitemShut {NoStop}%
\bibitem [{\citenamefont {Dusuel}\ \emph {et~al.}(2011)\citenamefont {Dusuel},
  \citenamefont {Kamfor}, \citenamefont {Or\'{u}s}, \citenamefont {Schmidt},\
  and\ \citenamefont {Vidal}}]{Dusuel_2011}%
  \BibitemOpen
  \bibfield  {author} {\bibinfo {author} {\bibfnamefont {S.}~\bibnamefont
  {Dusuel}}, \bibinfo {author} {\bibfnamefont {M.}~\bibnamefont {Kamfor}},
  \bibinfo {author} {\bibfnamefont {R.}~\bibnamefont {Or\'{u}s}}, \bibinfo
  {author} {\bibfnamefont {K.~P.}\ \bibnamefont {Schmidt}}, \ and\ \bibinfo
  {author} {\bibfnamefont {J.}~\bibnamefont {Vidal}},\ }\href {\doibase
  10.1103/PhysRevLett.106.107203} {\bibfield  {journal} {\bibinfo  {journal}
  {Phys. Rev. Lett.}\ }\textbf {\bibinfo {volume} {106}},\ \bibinfo {pages}
  {107203} (\bibinfo {year} {2011})}\BibitemShut {NoStop}%
\bibitem [{\citenamefont {Schmidt}(2013)}]{Schmidt_2013}%
  \BibitemOpen
  \bibfield  {author} {\bibinfo {author} {\bibfnamefont {K.~P.}\ \bibnamefont
  {Schmidt}},\ }\href {\doibase 10.1103/PhysRevB.88.035118} {\bibfield
  {journal} {\bibinfo  {journal} {Phys. Rev. B}\ }\textbf {\bibinfo {volume}
  {88}},\ \bibinfo {pages} {035118} (\bibinfo {year} {2013})}\BibitemShut
  {NoStop}%
\bibitem [{\citenamefont {Jahromi}\ \emph {et~al.}(2013)\citenamefont
  {Jahromi}, \citenamefont {Kargarian}, \citenamefont {Masoudi},\ and\
  \citenamefont {Schmidt}}]{Jahromi_2013}%
  \BibitemOpen
  \bibfield  {author} {\bibinfo {author} {\bibfnamefont {S.~S.}\ \bibnamefont
  {Jahromi}}, \bibinfo {author} {\bibfnamefont {M.}~\bibnamefont {Kargarian}},
  \bibinfo {author} {\bibfnamefont {S.~F.}\ \bibnamefont {Masoudi}}, \ and\
  \bibinfo {author} {\bibfnamefont {K.~P.}\ \bibnamefont {Schmidt}},\ }\href
  {\doibase 10.1103/PhysRevB.87.094413} {\bibfield  {journal} {\bibinfo
  {journal} {Phys. Rev. B}\ }\textbf {\bibinfo {volume} {87}},\ \bibinfo
  {pages} {094413} (\bibinfo {year} {2013})}\BibitemShut {NoStop}%
\bibitem [{\citenamefont {Morampudi}\ \emph {et~al.}(2014)\citenamefont
  {Morampudi}, \citenamefont {von Keyserlingk},\ and\ \citenamefont
  {Pollmann}}]{Morampudi_2014}%
  \BibitemOpen
  \bibfield  {author} {\bibinfo {author} {\bibfnamefont {S.~C.}\ \bibnamefont
  {Morampudi}}, \bibinfo {author} {\bibfnamefont {C.}~\bibnamefont {von
  Keyserlingk}}, \ and\ \bibinfo {author} {\bibfnamefont {F.}~\bibnamefont
  {Pollmann}},\ }\href {\doibase 10.1103/PhysRevB.90.035117} {\bibfield
  {journal} {\bibinfo  {journal} {Phys. Rev. B}\ }\textbf {\bibinfo {volume}
  {90}},\ \bibinfo {pages} {035117} (\bibinfo {year} {2014})}\BibitemShut
  {NoStop}%
\bibitem [{\citenamefont {Schulz}\ and\ \citenamefont
  {Burnell}(2016)}]{Schulz_2016}%
  \BibitemOpen
  \bibfield  {author} {\bibinfo {author} {\bibfnamefont {M.~D.}\ \bibnamefont
  {Schulz}}\ and\ \bibinfo {author} {\bibfnamefont {F.~J.}\ \bibnamefont
  {Burnell}},\ }\href {\doibase 10.1103/PhysRevB.94.165110} {\bibfield
  {journal} {\bibinfo  {journal} {Phys. Rev. B}\ }\textbf {\bibinfo {volume}
  {94}},\ \bibinfo {pages} {165110} (\bibinfo {year} {2016})}\BibitemShut
  {NoStop}%
\bibitem [{\citenamefont {Zhang}\ \emph {et~al.}(2017)\citenamefont {Zhang},
  \citenamefont {Melko},\ and\ \citenamefont {Kim}}]{Zhang_2017}%
  \BibitemOpen
  \bibfield  {author} {\bibinfo {author} {\bibfnamefont {Y.}~\bibnamefont
  {Zhang}}, \bibinfo {author} {\bibfnamefont {R.~G.}\ \bibnamefont {Melko}}, \
  and\ \bibinfo {author} {\bibfnamefont {E.-A.}\ \bibnamefont {Kim}},\ }\href
  {\doibase 10.1103/PhysRevB.96.245119} {\bibfield  {journal} {\bibinfo
  {journal} {Phys. Rev. B}\ }\textbf {\bibinfo {volume} {96}},\ \bibinfo
  {pages} {245119} (\bibinfo {year} {2017})}\BibitemShut {NoStop}%
\bibitem [{\citenamefont {Vanderstraeten}\ \emph {et~al.}(2017)\citenamefont
  {Vanderstraeten}, \citenamefont {Mari\"en}, \citenamefont {Haegeman},
  \citenamefont {Schuch}, \citenamefont {Vidal},\ and\ \citenamefont
  {Verstraete}}]{Vanderstraeten_2017}%
  \BibitemOpen
  \bibfield  {author} {\bibinfo {author} {\bibfnamefont {L.}~\bibnamefont
  {Vanderstraeten}}, \bibinfo {author} {\bibfnamefont {M.}~\bibnamefont
  {Mari\"en}}, \bibinfo {author} {\bibfnamefont {J.}~\bibnamefont {Haegeman}},
  \bibinfo {author} {\bibfnamefont {N.}~\bibnamefont {Schuch}}, \bibinfo
  {author} {\bibfnamefont {J.}~\bibnamefont {Vidal}}, \ and\ \bibinfo {author}
  {\bibfnamefont {F.}~\bibnamefont {Verstraete}},\ }\href {\doibase
  10.1103/PhysRevLett.119.070401} {\bibfield  {journal} {\bibinfo  {journal}
  {Phys. Rev. Lett.}\ }\textbf {\bibinfo {volume} {119}},\ \bibinfo {pages}
  {070401} (\bibinfo {year} {2017})}\BibitemShut {NoStop}%
\bibitem [{\citenamefont {Alicki}\ \emph {et~al.}(2009)\citenamefont {Alicki},
  \citenamefont {Fannes},\ and\ \citenamefont {Horodecki}}]{Alicki_2009}%
  \BibitemOpen
  \bibfield  {author} {\bibinfo {author} {\bibfnamefont {R.}~\bibnamefont
  {Alicki}}, \bibinfo {author} {\bibfnamefont {M.}~\bibnamefont {Fannes}}, \
  and\ \bibinfo {author} {\bibfnamefont {M.}~\bibnamefont {Horodecki}},\ }\href
  {\doibase 10.1088/1751-8113/42/6/065303} {\bibfield  {journal} {\bibinfo
  {journal} {J. Phys. A: Math. Theor.}\ }\textbf {\bibinfo {volume} {42}},\
  \bibinfo {pages} {065303} (\bibinfo {year} {2009})}\BibitemShut {NoStop}%
\bibitem [{\citenamefont {Castelnovo}\ and\ \citenamefont
  {Chamon}(2007)}]{Castelnovo_2007}%
  \BibitemOpen
  \bibfield  {author} {\bibinfo {author} {\bibfnamefont {C.}~\bibnamefont
  {Castelnovo}}\ and\ \bibinfo {author} {\bibfnamefont {C.}~\bibnamefont
  {Chamon}},\ }\href {\doibase 10.1103/PhysRevB.76.184442} {\bibfield
  {journal} {\bibinfo  {journal} {Phys. Rev. B}\ }\textbf {\bibinfo {volume}
  {76}},\ \bibinfo {pages} {184442} (\bibinfo {year} {2007})}\BibitemShut
  {NoStop}%
\bibitem [{\citenamefont {Nussinov}\ and\ \citenamefont
  {Ortiz}(2009)}]{Nussinov_2009_b}%
  \BibitemOpen
  \bibfield  {author} {\bibinfo {author} {\bibfnamefont {Z.}~\bibnamefont
  {Nussinov}}\ and\ \bibinfo {author} {\bibfnamefont {G.}~\bibnamefont
  {Ortiz}},\ }\href {\doibase 10.1073/pnas.0803726105} {\bibfield  {journal}
  {\bibinfo  {journal} {Proc. Nat. Acad. Sci.}\ }\textbf {\bibinfo {volume}
  {106}},\ \bibinfo {pages} {16944} (\bibinfo {year} {2009})}\BibitemShut
  {NoStop}%
\bibitem [{\citenamefont {Hal\'asz}\ and\ \citenamefont
  {Hamma}(2012)}]{Halasz_2012}%
  \BibitemOpen
  \bibfield  {author} {\bibinfo {author} {\bibfnamefont {G.~B.}\ \bibnamefont
  {Hal\'asz}}\ and\ \bibinfo {author} {\bibfnamefont {A.}~\bibnamefont
  {Hamma}},\ }\href {\doibase 10.1103/PhysRevA.86.062330} {\bibfield  {journal}
  {\bibinfo  {journal} {Phys. Rev. A}\ }\textbf {\bibinfo {volume} {86}},\
  \bibinfo {pages} {062330} (\bibinfo {year} {2012})}\BibitemShut {NoStop}%
\bibitem [{\citenamefont {Santra}\ \emph {et~al.}(2014)\citenamefont {Santra},
  \citenamefont {Hamma}, \citenamefont {Cincio}, \citenamefont {Subasi},
  \citenamefont {Zanardi},\ and\ \citenamefont {Amico}}]{Santra_2014}%
  \BibitemOpen
  \bibfield  {author} {\bibinfo {author} {\bibfnamefont {S.}~\bibnamefont
  {Santra}}, \bibinfo {author} {\bibfnamefont {A.}~\bibnamefont {Hamma}},
  \bibinfo {author} {\bibfnamefont {L.}~\bibnamefont {Cincio}}, \bibinfo
  {author} {\bibfnamefont {Y.}~\bibnamefont {Subasi}}, \bibinfo {author}
  {\bibfnamefont {P.}~\bibnamefont {Zanardi}}, \ and\ \bibinfo {author}
  {\bibfnamefont {L.}~\bibnamefont {Amico}},\ }\href {\doibase
  10.1103/PhysRevB.90.245128} {\bibfield  {journal} {\bibinfo  {journal} {Phys.
  Rev. B}\ }\textbf {\bibinfo {volume} {90}},\ \bibinfo {pages} {245128}
  (\bibinfo {year} {2014})}\BibitemShut {NoStop}%
\bibitem [{\citenamefont {Kamfor}\ \emph {et~al.}(2014)\citenamefont {Kamfor},
  \citenamefont {Dusuel}, \citenamefont {Vidal},\ and\ \citenamefont
  {Schmidt}}]{Kamfor_2014}%
  \BibitemOpen
  \bibfield  {author} {\bibinfo {author} {\bibfnamefont {M.}~\bibnamefont
  {Kamfor}}, \bibinfo {author} {\bibfnamefont {S.}~\bibnamefont {Dusuel}},
  \bibinfo {author} {\bibfnamefont {J.}~\bibnamefont {Vidal}}, \ and\ \bibinfo
  {author} {\bibfnamefont {K.~P.}\ \bibnamefont {Schmidt}},\ }\href {\doibase
  10.1103/PhysRevB.89.045411} {\bibfield  {journal} {\bibinfo  {journal} {Phys.
  Rev. B}\ }\textbf {\bibinfo {volume} {89}},\ \bibinfo {pages} {045411}
  (\bibinfo {year} {2014})}\BibitemShut {NoStop}%
\bibitem [{\citenamefont {Hamma}\ \emph {et~al.}(2005)\citenamefont {Hamma},
  \citenamefont {Zanardi},\ and\ \citenamefont {Wen}}]{Hamma_2005}%
  \BibitemOpen
  \bibfield  {author} {\bibinfo {author} {\bibfnamefont {A.}~\bibnamefont
  {Hamma}}, \bibinfo {author} {\bibfnamefont {P.}~\bibnamefont {Zanardi}}, \
  and\ \bibinfo {author} {\bibfnamefont {X.-G.}\ \bibnamefont {Wen}},\ }\href
  {\doibase 10.1103/PhysRevB.72.035307} {\bibfield  {journal} {\bibinfo
  {journal} {Phys. Rev. B}\ }\textbf {\bibinfo {volume} {72}},\ \bibinfo
  {pages} {035307} (\bibinfo {year} {2005})}\BibitemShut {NoStop}%
\bibitem [{\citenamefont {Nussinov}\ and\ \citenamefont
  {Ortiz}(2008)}]{Nussinov_2008}%
  \BibitemOpen
  \bibfield  {author} {\bibinfo {author} {\bibfnamefont {Z.}~\bibnamefont
  {Nussinov}}\ and\ \bibinfo {author} {\bibfnamefont {G.}~\bibnamefont
  {Ortiz}},\ }\href {\doibase 10.1103/PhysRevB.77.064302} {\bibfield  {journal}
  {\bibinfo  {journal} {Phys. Rev. B}\ }\textbf {\bibinfo {volume} {77}},\
  \bibinfo {pages} {064302} (\bibinfo {year} {2008})}\BibitemShut {NoStop}%
\bibitem [{\citenamefont {Reiss}\ and\ \citenamefont
  {Schmidt}(2019)}]{Reiss_2019}%
  \BibitemOpen
  \bibfield  {author} {\bibinfo {author} {\bibfnamefont {D.~A.}\ \bibnamefont
  {Reiss}}\ and\ \bibinfo {author} {\bibfnamefont {K.~P.}\ \bibnamefont
  {Schmidt}},\ }\href {\doibase 10.21468/SciPostPhys.6.6.078} {\bibfield
  {journal} {\bibinfo  {journal} {SciPost Phys.}\ }\textbf {\bibinfo {volume}
  {6}},\ \bibinfo {pages} {78} (\bibinfo {year} {2019})}\BibitemShut {NoStop}%
\bibitem [{\citenamefont {Wen}(2000)}]{Wen_2010}%
  \BibitemOpen
  \bibfield  {author} {\bibinfo {author} {\bibfnamefont {X.-G.}\ \bibnamefont
  {Wen}},\ }\href {\doibase 10.1103/PhysRevLett.84.3950} {\bibfield  {journal}
  {\bibinfo  {journal} {Phys. Rev. Lett.}\ }\textbf {\bibinfo {volume} {84}},\
  \bibinfo {pages} {3950} (\bibinfo {year} {2000})}\BibitemShut {NoStop}%
\bibitem [{\citenamefont {Barkeshli}\ and\ \citenamefont
  {Wen}(2010)}]{Barkeshli_2010}%
  \BibitemOpen
  \bibfield  {author} {\bibinfo {author} {\bibfnamefont {M.}~\bibnamefont
  {Barkeshli}}\ and\ \bibinfo {author} {\bibfnamefont {X.-G.}\ \bibnamefont
  {Wen}},\ }\href {\doibase 10.1103/PhysRevLett.105.216804} {\bibfield
  {journal} {\bibinfo  {journal} {Phys. Rev. Lett.}\ }\textbf {\bibinfo
  {volume} {105}},\ \bibinfo {pages} {216804} (\bibinfo {year}
  {2010})}\BibitemShut {NoStop}%
\bibitem [{\citenamefont {M\"oller}\ \emph {et~al.}(2014)\citenamefont
  {M\"oller}, \citenamefont {Hormozi}, \citenamefont {Slingerland},\ and\
  \citenamefont {Simon}}]{Moeller_2014}%
  \BibitemOpen
  \bibfield  {author} {\bibinfo {author} {\bibfnamefont {G.}~\bibnamefont
  {M\"oller}}, \bibinfo {author} {\bibfnamefont {L.}~\bibnamefont {Hormozi}},
  \bibinfo {author} {\bibfnamefont {J.}~\bibnamefont {Slingerland}}, \ and\
  \bibinfo {author} {\bibfnamefont {S.~H.}\ \bibnamefont {Simon}},\ }\href
  {\doibase 10.1103/PhysRevB.90.235101} {\bibfield  {journal} {\bibinfo
  {journal} {Phys. Rev. B}\ }\textbf {\bibinfo {volume} {90}},\ \bibinfo
  {pages} {235101} (\bibinfo {year} {2014})}\BibitemShut {NoStop}%
\bibitem [{\citenamefont {Bombin}\ and\ \citenamefont
  {Martin-Delgado}(2008)}]{Bombin_2008}%
  \BibitemOpen
  \bibfield  {author} {\bibinfo {author} {\bibfnamefont {H.}~\bibnamefont
  {Bombin}}\ and\ \bibinfo {author} {\bibfnamefont {M.~A.}\ \bibnamefont
  {Martin-Delgado}},\ }\href {\doibase 10.1103/PhysRevB.78.115421} {\bibfield
  {journal} {\bibinfo  {journal} {Phys. Rev. B}\ }\textbf {\bibinfo {volume}
  {78}},\ \bibinfo {pages} {115421} (\bibinfo {year} {2008})}\BibitemShut
  {NoStop}%
\bibitem [{\citenamefont {Vidal}\ \emph {et~al.}(2008)\citenamefont {Vidal},
  \citenamefont {Schmidt},\ and\ \citenamefont {Dusuel}}]{Vidal_2008}%
  \BibitemOpen
  \bibfield  {author} {\bibinfo {author} {\bibfnamefont {J.}~\bibnamefont
  {Vidal}}, \bibinfo {author} {\bibfnamefont {K.~P.}\ \bibnamefont {Schmidt}},
  \ and\ \bibinfo {author} {\bibfnamefont {S.}~\bibnamefont {Dusuel}},\ }\href
  {\doibase 10.1103/PhysRevB.78.245121} {\bibfield  {journal} {\bibinfo
  {journal} {Phys. Rev. B}\ }\textbf {\bibinfo {volume} {78}},\ \bibinfo
  {pages} {245121} (\bibinfo {year} {2008})}\BibitemShut {NoStop}%
\bibitem [{\citenamefont {Castelnovo}\ and\ \citenamefont
  {Chamon}(2008)}]{Castelnovo_2008}%
  \BibitemOpen
  \bibfield  {author} {\bibinfo {author} {\bibfnamefont {C.}~\bibnamefont
  {Castelnovo}}\ and\ \bibinfo {author} {\bibfnamefont {C.}~\bibnamefont
  {Chamon}},\ }\href {\doibase 10.1103/PhysRevB.78.155120} {\bibfield
  {journal} {\bibinfo  {journal} {Phys. Rev. B}\ }\textbf {\bibinfo {volume}
  {78}},\ \bibinfo {pages} {155120} (\bibinfo {year} {2008})}\BibitemShut
  {NoStop}%
\bibitem [{\citenamefont {Schmidt}\ \emph {et~al.}(2008)\citenamefont
  {Schmidt}, \citenamefont {Dusuel},\ and\ \citenamefont
  {Vidal}}]{Schmidt_2008}%
  \BibitemOpen
  \bibfield  {author} {\bibinfo {author} {\bibfnamefont {K.~P.}\ \bibnamefont
  {Schmidt}}, \bibinfo {author} {\bibfnamefont {S.}~\bibnamefont {Dusuel}}, \
  and\ \bibinfo {author} {\bibfnamefont {J.}~\bibnamefont {Vidal}},\ }\href
  {\doibase 10.1103/PhysRevLett.100.057208} {\bibfield  {journal} {\bibinfo
  {journal} {Phys. Rev. Lett.}\ }\textbf {\bibinfo {volume} {100}},\ \bibinfo
  {pages} {057208} (\bibinfo {year} {2008})}\BibitemShut {NoStop}%
\bibitem [{\citenamefont {He}\ \emph {et~al.}(1990)\citenamefont {He},
  \citenamefont {Hamer},\ and\ \citenamefont {Oitmaa}}]{He_1990}%
  \BibitemOpen
  \bibfield  {author} {\bibinfo {author} {\bibfnamefont {H.~X.}\ \bibnamefont
  {He}}, \bibinfo {author} {\bibfnamefont {C.~J.}\ \bibnamefont {Hamer}}, \
  and\ \bibinfo {author} {\bibfnamefont {J.}~\bibnamefont {Oitmaa}},\ }\href
  {\doibase 10.1088/0305-4470/23/10/018} {\bibfield  {journal} {\bibinfo
  {journal} {Journal of Physics A: Mathematical and General}\ }\textbf
  {\bibinfo {volume} {23}},\ \bibinfo {pages} {1775} (\bibinfo {year}
  {1990})}\BibitemShut {NoStop}%
\bibitem [{\citenamefont {Bl\"ote}\ and\ \citenamefont
  {Deng}(2002)}]{Bloete_2002}%
  \BibitemOpen
  \bibfield  {author} {\bibinfo {author} {\bibfnamefont {H.~W.~J.}\
  \bibnamefont {Bl\"ote}}\ and\ \bibinfo {author} {\bibfnamefont
  {Y.}~\bibnamefont {Deng}},\ }\href {\doibase 10.1103/PhysRevE.66.066110}
  {\bibfield  {journal} {\bibinfo  {journal} {Phys. Rev. E 􏰒􏰀}\ }\textbf
  {\bibinfo {volume} {66}},\ \bibinfo {pages} {066110} (\bibinfo {year}
  {2002})}\BibitemShut {NoStop}%
\bibitem [{\citenamefont {Schuler}\ \emph {et~al.}(2016)\citenamefont
  {Schuler}, \citenamefont {Whitsitt}, \citenamefont {Henry}, \citenamefont
  {Sachdev},\ and\ \citenamefont {L\"auchli}}]{Schuler_2016}%
  \BibitemOpen
  \bibfield  {author} {\bibinfo {author} {\bibfnamefont {M.}~\bibnamefont
  {Schuler}}, \bibinfo {author} {\bibfnamefont {S.}~\bibnamefont {Whitsitt}},
  \bibinfo {author} {\bibfnamefont {L.-P.}\ \bibnamefont {Henry}}, \bibinfo
  {author} {\bibfnamefont {S.}~\bibnamefont {Sachdev}}, \ and\ \bibinfo
  {author} {\bibfnamefont {A.~M.}\ \bibnamefont {L\"auchli}},\ }\href {\doibase
  10.1103/PhysRevLett.117.210401} {\bibfield  {journal} {\bibinfo  {journal}
  {Phys. Rev. Lett.}\ }\textbf {\bibinfo {volume} {117}},\ \bibinfo {pages}
  {210401} (\bibinfo {year} {2016})}\BibitemShut {NoStop}%
\bibitem [{\citenamefont {Levin}\ and\ \citenamefont {Wen}(2006)}]{Levin_2006}%
  \BibitemOpen
  \bibfield  {author} {\bibinfo {author} {\bibfnamefont {M.~A.}\ \bibnamefont
  {Levin}}\ and\ \bibinfo {author} {\bibfnamefont {X.-G.}\ \bibnamefont
  {Wen}},\ }\href {\doibase 10.1103/PhysRevLett.96.110405} {\bibfield
  {journal} {\bibinfo  {journal} {Phys. Rev. Lett.}\ }\textbf {\bibinfo
  {volume} {96}},\ \bibinfo {pages} {110405} (\bibinfo {year}
  {2006})}\BibitemShut {NoStop}%
\bibitem [{\citenamefont {Kitaev}\ and\ \citenamefont
  {Preskill}(2006)}]{Kitaev_2006_b}%
  \BibitemOpen
  \bibfield  {author} {\bibinfo {author} {\bibfnamefont {A.}~\bibnamefont
  {Kitaev}}\ and\ \bibinfo {author} {\bibfnamefont {J.}~\bibnamefont
  {Preskill}},\ }\href {\doibase 10.1103/PhysRevLett.96.110404} {\bibfield
  {journal} {\bibinfo  {journal} {Phys. Rev. Lett.}\ }\textbf {\bibinfo
  {volume} {96}},\ \bibinfo {pages} {110404} (\bibinfo {year}
  {2006})}\BibitemShut {NoStop}%
\bibitem [{\citenamefont {Kitaev}(2006)}]{Kitaev_2006}%
  \BibitemOpen
  \bibfield  {author} {\bibinfo {author} {\bibfnamefont {A.}~\bibnamefont
  {Kitaev}},\ }\href {\doibase 10.1016/j.aop.2005.10.005} {\bibfield  {journal}
  {\bibinfo  {journal} {Ann. Phys.}\ }\textbf {\bibinfo {volume} {321}},\
  \bibinfo {pages} {2} (\bibinfo {year} {2006})}\BibitemShut {NoStop}%
\bibitem [{\citenamefont {Seifert}\ \emph {et~al.}(2018)\citenamefont
  {Seifert}, \citenamefont {Gritsch}, \citenamefont {Wagner}, \citenamefont
  {Joshi}, \citenamefont {Brenig}, \citenamefont {Vojta},\ and\ \citenamefont
  {Schmidt}}]{Seifert_2018}%
  \BibitemOpen
  \bibfield  {author} {\bibinfo {author} {\bibfnamefont {U.~F.~P.}\
  \bibnamefont {Seifert}}, \bibinfo {author} {\bibfnamefont {J.}~\bibnamefont
  {Gritsch}}, \bibinfo {author} {\bibfnamefont {E.}~\bibnamefont {Wagner}},
  \bibinfo {author} {\bibfnamefont {D.~G.}\ \bibnamefont {Joshi}}, \bibinfo
  {author} {\bibfnamefont {W.}~\bibnamefont {Brenig}}, \bibinfo {author}
  {\bibfnamefont {M.}~\bibnamefont {Vojta}}, \ and\ \bibinfo {author}
  {\bibfnamefont {K.~P.}\ \bibnamefont {Schmidt}},\ }\href {\doibase
  10.1103/PhysRevB.98.155101} {\bibfield  {journal} {\bibinfo  {journal} {Phys.
  Rev. B}\ }\textbf {\bibinfo {volume} {98}},\ \bibinfo {pages} {155101}
  (\bibinfo {year} {2018})}\BibitemShut {NoStop}%
\end{thebibliography}%
\end{document}